%% file: UserBoost_manuscript.tex
\documentclass[sigconf,anonymous=false]{acmart}

\setcopyright{none}
\renewcommand\footnotetextcopyrightpermission[1]{}
\usepackage{graphicx}
\usepackage{multirow}
\usepackage{tabularx}
\usepackage{subcaption}
\usepackage{amsmath}
\usepackage{bm}
\usepackage{hyperref}
\usepackage{mathtools}
\usepackage{isomath}

\newcolumntype{Y}{>{\centering\arraybackslash}X}

\newcommand{\paraHeading}[1]{\noindent{\textbf{#1}.}}

\AtBeginDocument{%
  }

\begin{document}

\title{UserBoost: Generating User-specific Synthetic Data for Faster Enrolment into Behavioural Biometric Systems}

\author{George Webber}
\orcid{0009-0003-2573-7690}
\affiliation{%
  \institution{King's College London}
  \city{London}
  \country{UK}
}
\email{george.webber@kcl.ac.uk}

\author{Jack Sturgess}
\orcid{0000-0001-5708-3052}
\affiliation{%
  \institution{University of Oxford}
  \city{Oxford}
  \country{UK}
}
\email{jack.sturgess@cs.ox.ac.uk}

\author{Ivan Martinovic}
\orcid{0000-0003-2340-3040}
\affiliation{%
  \institution{University of Oxford}
  \city{Oxford}
  \country{UK}
}
\email{ivan.martinovic@cs.ox.ac.uk}

\renewcommand{\shortauthors}{Webber et al.}

\begin{abstract}
Behavioural biometric authentication systems entail an enrolment period that is burdensome for the user.
In this work, we explore generating synthetic gestures from a few real user gestures with generative deep learning, with the application of training a simple (i.e. non-deep-learned) authentication model. Specifically, we show that utilising synthetic data alongside real data can reduce the number of real datapoints a user must provide to enrol into a biometric system.
To validate our methods, we use the publicly available dataset of WatchAuth, a system proposed in 2022 for authenticating smartwatch payments using the physical gesture of reaching towards a payment terminal.
We develop a regularised autoencoder model for generating synthetic user-specific wrist motion data representing these physical gestures, and demonstrate the diversity and fidelity of our synthetic gestures.
We show that using synthetic gestures in training can improve classification ability for a real-world system.
Through this technique we can reduce the number of gestures required to enrol a user into a WatchAuth-like system by more than 40\% without negatively impacting its error rates. 
\end{abstract}




\maketitle

\section{Introduction}

Many biometric authentication systems require a burdensome enrolment period for the user, where they must provide many instances of their biometric signal. This reduces the usability of the authentication system. Ideally, a user would be able to enrol into the security system with fewer instances of their biometric signal.

Generating synthetic data to supplement small, real-world datasets has potential cost, privacy and speed advantages for many tasks when compared to collecting large datasets from real-world users.
While used widely in such computer science domains as healthcare technology and image analysis, synthetic data remains relatively underutilised in security.

In this work, we investigate generating synthetic gesture data for a smartwatch. A gesture is a pattern of user movements and is measured over time by the Inertial Measurement Unit (IMU) on the device. We focus on smartwatch authentication for this work because smartwatch authentication models must be compact to satisfy hardware requirements. As a result, state-of-the-art deep learning models that could improve the authentication system are too memory-intensive to install on the smartwatch. However, given a corpus of user data, a deep learning model could teach a simpler, smartwatch-compatible authentication model by generating synthetic gesture data. We investigate the specific use case for this data in improving the training of a motion-based authentication system for smartwatch contactless payments.

Making contactless payments using near-field communication (NFC) is commonplace in today's society.
Wearable devices such as smartwatches are an increasingly popular medium for utilising contactless payment technology --- US payments made by smartwatch are projected to increase from $\sim$US\$40 billion in 2020 to $\sim$US\$500 billion by 2024~\cite{PayingYourWrist2023}. This rise in adoption drives a need for biometric smartwatch authentication to prevent fraudulent users making payments.
Smartwatches typically lack sophisticated fingerprint sensors or cameras, but usually include an accelerometer and gyroscope for motion sensing.
In 2022, Sturgess et al. introduced WatchAuth~\cite{sturgessWatchAuthUserAuthentication2022}, a machine learning system for authenticating a user's payment based on the physical act of extending the smartwatch to a payment terminal, a motion we call a \textit{payment gesture} (see Figure \ref{fig:combined_intro_images}).

The attractiveness of smartwatch payments lies mostly in convenience, and so WatchAuth was conceived to improve the security of a payment without negatively affecting usability.
However, similar to many biometric authentication systems, it requires a burdensome enrolment phase to become a useful security measure.
In this phase, a user must provide many example gestures (on the order of 100) to create a unique template, which the system then compares to future gestures for authentication.
Reducing the burden of this enrolment phase would encourage users to use the additional security on offer.

We consider the application of generative deep learning to produce synthetic data with a view to reducing the enrolment burden.

This work makes the following contributions:
\begin{itemize}
    \item We formulate and present a novel use case for reducing the enrolment burden for an authentication system through the application of synthetic data.
    \item We propose an autoencoder-based model that attempts to model the distribution of payment gestures, and can be used to generate synthetic gestures.
    \item We show that on a real-world dataset with a limited number of users (n=16), our system can reduce WatchAuth enrolment effort by more than 40\%, without negatively affecting its security or usability.
    \item We make available our codebase for training generative models, as well as for formatting the WatchAuth dataset for deep learning applications at\\https://github.com/GeorgeWebber/WatchAuth-DL.
\end{itemize}

\section{Related Work}\label{ch:3-related work}

The general problem of data scarcity for classification tasks on IMU data is well known, and many attempts to solve it in different contexts with synthetic data have been made.
Kim \& Brown ~\cite{kimImprovingAmericanSign2019} apply pre-defined transformations to real data to generate additional training data for the task of American Sign Language recognition from IMU data.
Panneer et al.~\cite{santhalingamSyntheticSmartwatchIMU2023} take a different approach, presenting a video-to-IMU translation framework for converting a large number of publicly available videos to IMU training data.
Alzantot et al.~\cite{alzantotSenseGenDeepLearning2017} use deep learning for generating synthetic timeseries sensor data capable of fooling a discriminator network, aiming to replace personal data with synthetic data for increased privacy. However, they do not show the ability to generate user-specific samples.

In their review of timeseries data augmentation methods, Iglesias et al.~\cite{iglesiasDataAugmentationTechniques2023} identify VAEs and generative adversarial networks (GANs) as the primary methods for deep generative data augmentation. Particular works of relevance for this paper are Alawneh et al.~\cite{alawnehEnhancingHumanActivity2021}, who use VAEs for data augmentation in a human activity recognition context, and Goubeaud et al.~\cite{goubeaudUsingVariationalAutoencoder2021}, who show that data generated from a VAE improves the performance of random forest classifiers in contexts unrelated to security.

The related field of Human Activity Recognition (HAR) has explored the potential of deep-learning generation of synthetic data, and shown that this synthetic data can augment training of machine learning systems. Zhang et al.~\cite{zhangDeepLearningHuman2022} details a number of successful studies using GANs to augment training datasets for HAR. For example, de Souza et al.~\cite{desouzaExploringImpactSynthetic2023} show the feasibility of improving HAR performance from wearable data. Their study shows that wearable timeseries data can be generated that is representative, diverse, and accurate, and that training a machine learning system with this data can improve the quality of HAR classification. Similar conclusions are reached by other authors using adversarial methods, including Alharbi et al.~\cite{alharbiSyntheticSensorData2020} and Konak et al.~\cite{konakOvercomingDataScarcity}.

Zhang et al. also highlights the use case of transfer learning for HAR to improve performance across different subjects, sensors and sampling rates. In the most relevant work, Soleimani \& Nazerfard~\cite{soleimaniCrosssubjectTransferLearning2021} investigate the use of GANs to transfer user-specific characteristics onto data from a second user, and show that this augmented data can make an activity classifier more independent of the user. This work showcases a method for generating user-dependent synthetic data, but does not consider its usefulness in an authentication context.

\section{Background}\label{ch:2-background}

\subsection{Motivation}

Sturgess et al.~\cite{sturgessWatchAuthUserAuthentication2022} show that a payment gesture, performed when a user taps a smartwatch against an NFC-enabled payment terminal, is a biometric that can authenticate a user.
The machine learning system they describe for this purpose is called WatchAuth, and can be run on a smartwatch without modifying payment terminals or smartwatch hardware.

The system model of WatchAuth assumes that a user is wearing a smartwatch on their wrist and uses it to make an NFC-enabled payment at a point-of-sale terminal in a typical setting (for example, a shop or entry barriers to a transport system).
It is further assumed that the user performs a payment gesture by moving and rotating their wrist towards the terminal until it is near enough to exchange data via NFC ($<10$ cm), at which point the gesture ends and payment is either approved or denied.
WatchAuth only requires access to the watch's inertial sensors.
As it is a machine learning system, to achieve strong authentication performance it is trained on many positive and negative samples and can't be employed for user authentication until the user has exhibited sufficiently many payment gestures (i.e. positive samples), either naturally over a long time frame or as a burdensome initial enrolment phase.

A potential solution to the issue of a burdensome enrolment phase would be to generate user-specific synthetic data and augment the initial dataset with these synthetic positive instances to train WatchAuth.
Smartwatch-based models are necessarily resource-constrained.
This approach offers the benefit of using an unconstrained source of computing power to generate synthetic positive training examples, which are then used to train the smartwatch-compatible WatchAuth model.
The synthetic positive examples may themselves be generated by a machine learning model.
The approach offers the possibility of exploiting commonalities between the target user and a large corpus of seen users, which could include transferring user-specific characteristics of a payment gesture to a new position of payment terminal or leveraging characteristics common to age, height or sex.

\subsection{WatchAuth Dataset}

\begin{figure}[t]
	\centering
	\includegraphics[width=0.9\linewidth]{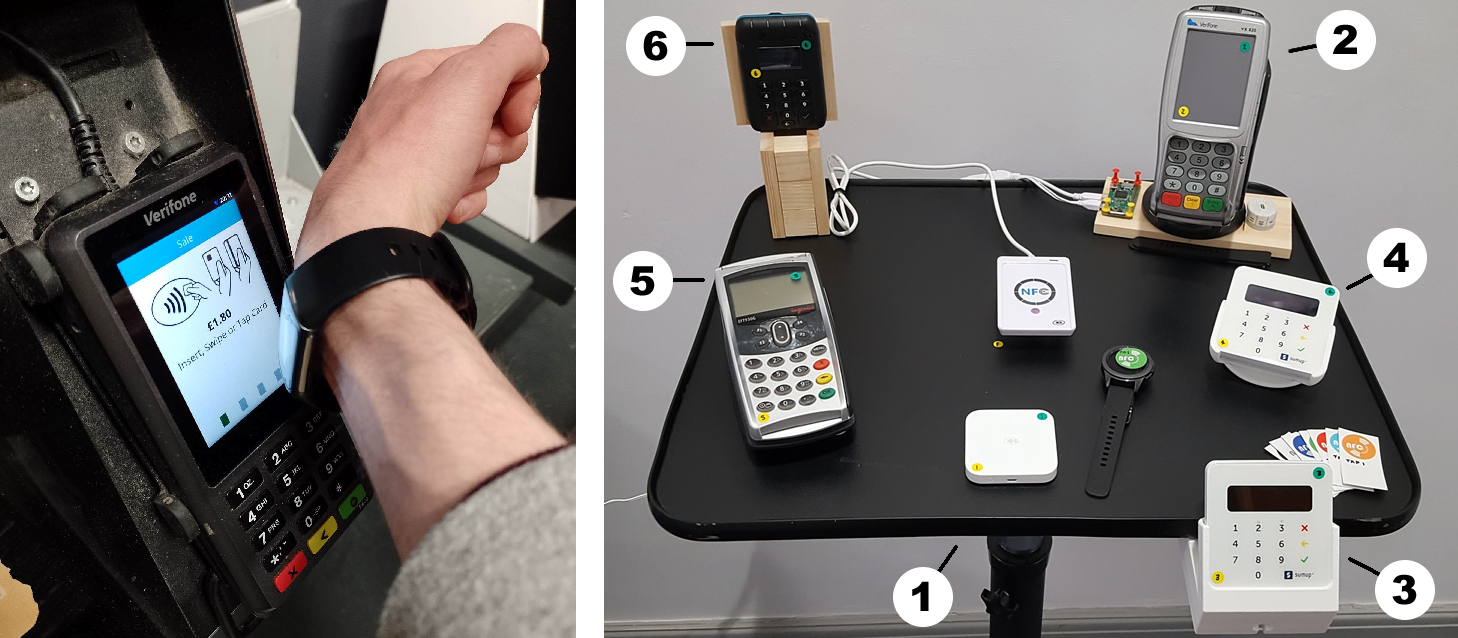}
	\caption[]%
	{{ \small Left: A smartwatch user making a \textit{payment gesture}. Right: The positioning of six contactless payment terminals used in WatchAuth experiments; the seventh terminal (centre) was handheld (image courtesy of~\cite{sturgessWatchAuthUserAuthentication2022}).}}
	\label{fig:combined_intro_images}
\end{figure}

All experiments in this paper make use of the publicly available WatchAuth dataset~\cite{sturgessWatchAuthDatasetExtended2023}.
This dataset contains data collected in a physical study with 16 participants.
Across three sessions, participants performed gestures simulating smartwatch payments at seven differently-positioned contactless payment terminals (see Figure \ref{fig:combined_intro_images}), with a watch on their wrist for data collection.

Data was sampled at 50 Hz from the watch's inbuilt sensors, returning vectors for acceleration (from an accelerometer), angular velocity (from a gyroscope), linear acceleration (from a linear accelerometer) and device orientation (derived from other sensors).
The dataset includes timestamps showing when contact between the watch and payment terminal was initiated (via NFC).

In addition to the authentication gesture data collected, the dataset also contains "non-gesture" data collected by users outside of the lab while performing everyday tasks. This data forms a useful negative class for gesture recognition.
In total, 3,484 gesture and 30,771 non-gesture datapoints were collected.

\begin{figure*}[t] 
	\centering
        \begin{subfigure}[b]{0.33\textwidth}
		\centering
		\includegraphics[width=\textwidth]{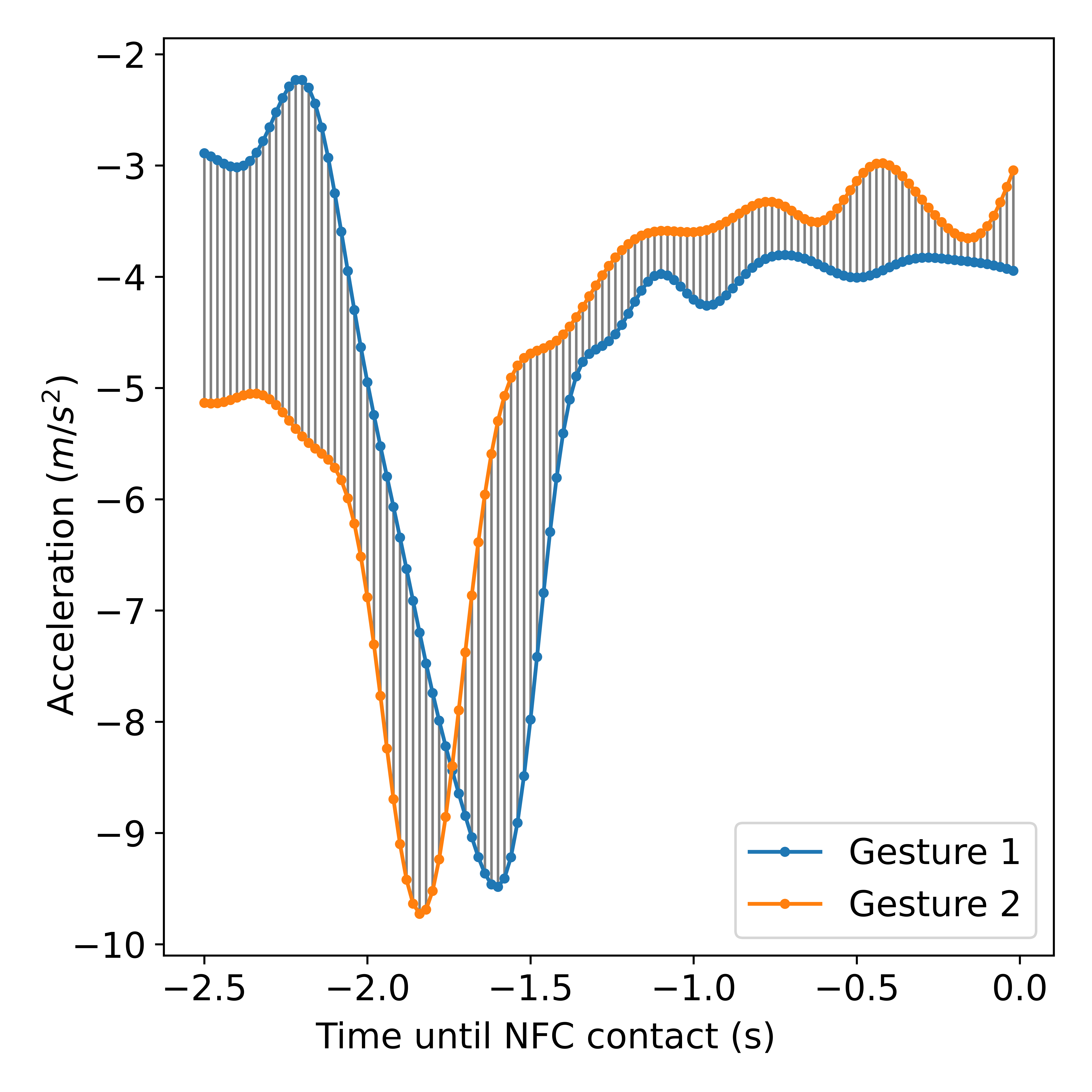}
		\caption[]%
		{{\small MSE}}
	\end{subfigure}
	\hfill
	\begin{subfigure}[b]{0.33\textwidth}
		\centering
		\includegraphics[width=\textwidth]{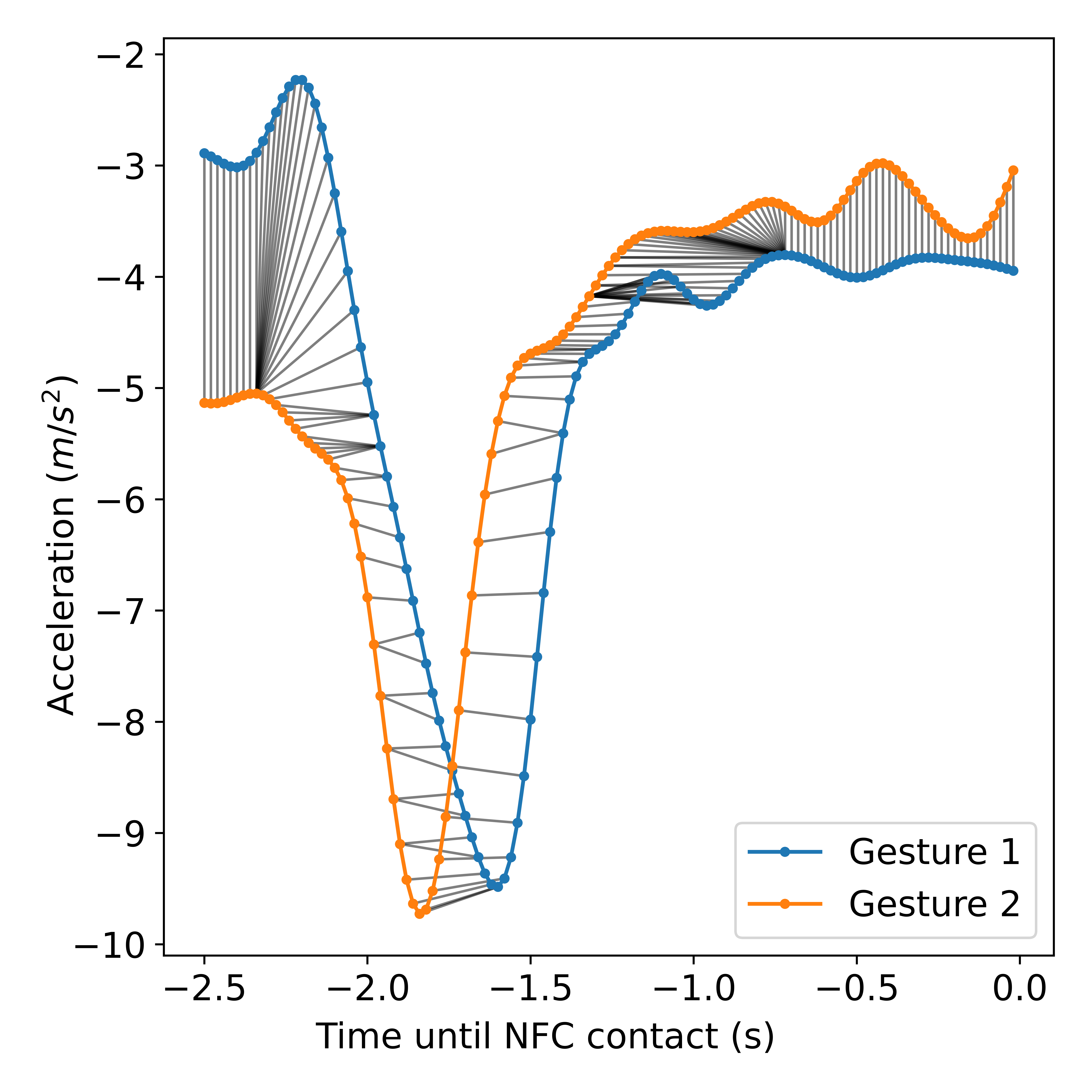}
		\caption[]%
		{{\small DTW}}
	\end{subfigure}
	\hfill
	\begin{subfigure}[b]{0.33\textwidth}
		\centering 
		\includegraphics[width=\textwidth]{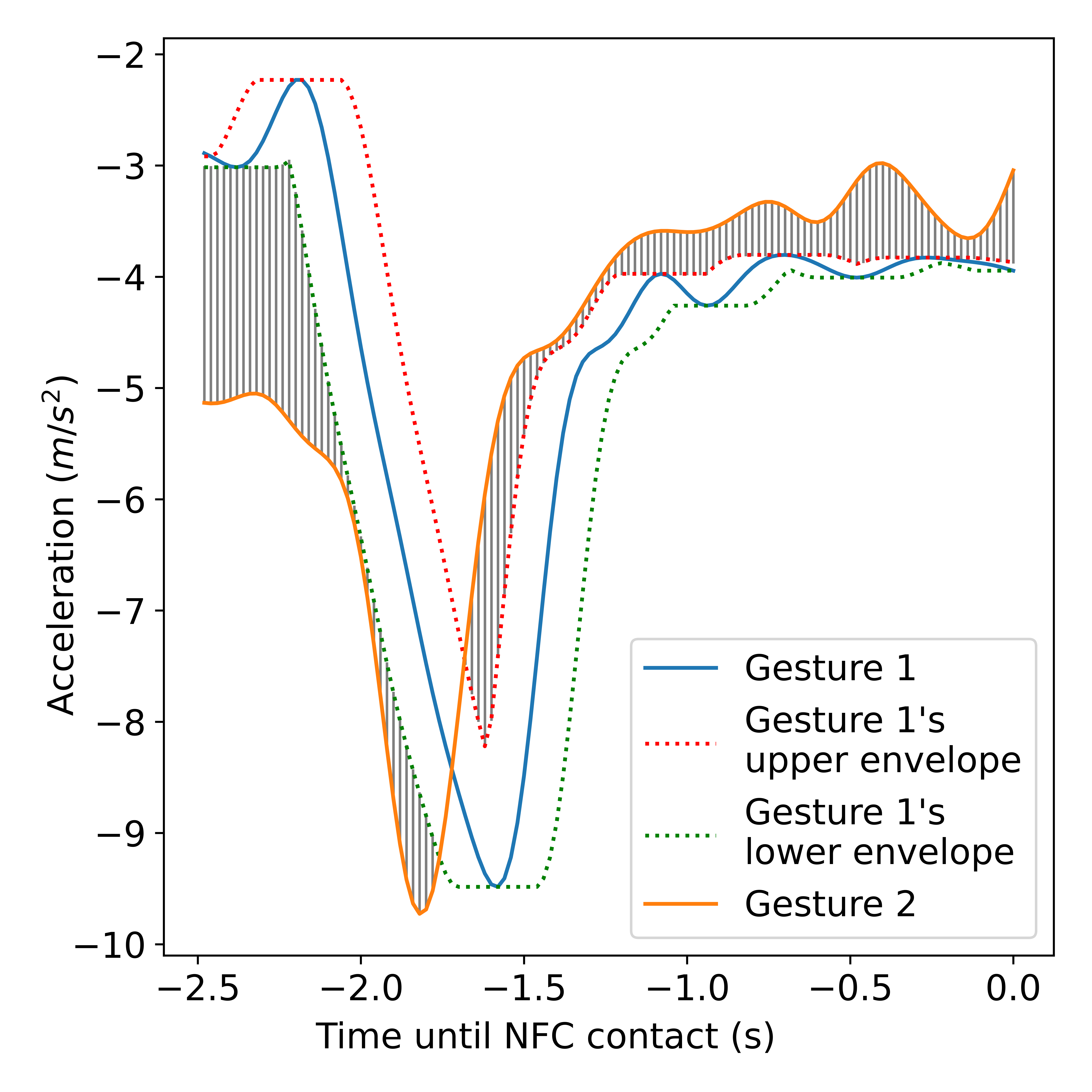}
		\caption[]%
		{{\small Keogh's lower bound}}
	\end{subfigure}
	\caption[]%
	{{\small The dissimilarity of two WatchAuth gestures, measured with (a) Mean Square Error (MSE) distance, (b) Dynamic Time Warping (DTW) distance, and (c) Keogh's lower bound. Matching between points is shown in grey.}}
	\label{fig:lbkeogh}
\end{figure*}

WatchAuth was designed to be memory-efficient and resource-sparse in order to work with the limited memory and battery capacity of a smartwatch.
To classify payment gestures, it first windows a gesture's data to the 4 seconds before NFC contact.
It then extracts manually-defined features (e.g. mean, variance, skew, peak count).
These features are calculated along each of the gesture data's channels (e.g. acceleration x-axis, gyroscope z-axis, acceleration Euclidean norm).
A random forest classifier with 100 decision trees is then trained to classify these features, using the target user's data as the positive class and other users' data as the negative class.

\subsection{Generative Modelling}

\paraHeading{Autoencoder models}
One common family of generative deep learning models are autoencoders.
Under this semi-supervised framework, we have two neural networks: an encoder $ E $ and decoder $ D $.
Given input $ x $, we first apply the encoder $ E $ to generate an embedding in a latent space (typically a Euclidean space of low dimension).
We then apply the decoder $ D $ to $ E(x)$ to reconstruct a signal in the original domain.
For a given loss function $ \ell$, the objective of the network is to minimise the reconstruction loss $ \ell(x, (D \circ E) (x)) $.

An autoencoder's latent space must be sufficiently regular to be useful for generating synthetic data.
In particular, we require it to be sufficiently continuous so that close together points decode to semantically similar representations~\cite{roccaUnderstandingVariationalAutoencoders2021}, and to be complete, i.e. without "holes".
Unregularised autoencoders only minimise reconstruction loss, which usually results in a latent space that does not encode useful semantic information and is full of holes where the decoder has never been trained~\cite{tolstikhinWassersteinAutoEncoders2018}.

The most well-known version of a regularised autoencoder is a variational autoencoder (VAE), proposed by Kingma \& Welling~\cite{kingmaAutoEncodingVariationalBayes2014a}.
Instead of outputting a single vector in the latent space, the encoder is modified to output a latent mean and variance representing the parameters of a Gaussian distribution.
The decoder then samples from this Gaussian and decodes the sampled point.

To comply with our latent space requirements of continuity and completeness, we want our latent variances to be sufficiently large and our latent means sufficiently close such that the latent distributions of different points overlap.
We achieve this using the Kullback-Leibler (KL) divergence to penalise the distance between each latent distribution and a standard Gaussian.
This gives rise to the following KL loss used as a regularisation term in the VAE loss function\footnote{In practice, our encoder outputs the \textit{logarithm} of the latent variance, resulting in a change of variables in this formula.}:
\begin{align*}
	\ell_{KL}(\mu, \sigma^2) &= D_{KL}(\mathcal{N}(0,1) \space || \space \mathcal{N}(\mu, \sigma^2)) \\
	\qquad &= -\frac{1}{2} (1 + log(\sigma^2) - \mu^2 - \sigma^2)
\end{align*}

The full loss function used in a VAE is the sum of the reconstruction and KL losses.

\paraHeading{Dynamic Time Warping (DTW)}
DTW is an algorithm for quantifying the dissimilarity between two timeseries that may not be perfectly aligned.
The DTW distance is computed as the minimum Euclidean distance between the timeseries across all temporal alignments.
A given temporal alignment may be described by a matching between two sequences (see Figure \ref{fig:lbkeogh}).

Given sequences $ \mathbf{x} = (x_i)_{i=1}^m $ and $ \mathbf{y} = (y_i)_{i=1}^n $, a valid matching must obey the following rules:

\begin{enumerate}
	\item The first and last points of the sequences must be matched to each other (but may also match other points).
	\item Every index from both sequences must be matched with at least one index from the other sequence.
	\item Indices from the first sequence must be matched in a monotonically increasing manner to indices from the second sequence, and vice versa.
\end{enumerate}

The cost of a matching is the sum of distances between matched points.
An optimal matching may be computed via dynamic programming in time and space complexity equal to the product of the lengths of the sequences.

\paraHeading{Keogh's lower bound for DTW}
Keogh's lower bound is a linear time lower bound on DTW that makes use of derived series, termed \textit{envelopes}, around a timeseries.
For a sequence $ \mathbf{x} = (x_i)_{i=1}^n $, its upper and lower envelopes $ U_x $ and $ L_x $ are the maximum and minimum values of $ \mathbf{x} $ within a moving window of half-width $ w $:
\begin{align*}
 \qquad U_i^{\mathbf{x}} &= \max_{\max\{1, i-w\} \le j \le \min \{ l, i + w \}} \{x_j\} \qquad \\
 \qquad L_i^{\mathbf{x}} &= \min_{\max\{1, i-w\} \le j \le \min \{ l, i + w \}} \{x_j\}
\end{align*}

Keogh's lower bound is the sum of distances at points where sequence $ \mathbf{y} = (y_i)_{i=1}^n $ is not contained within $ \mathbf{x} $'s envelopes (see Figure \ref{fig:lbkeogh}): 
\[
\qquad d_{KLB}^{(w)}(\mathbf{x}, \mathbf{y}) = \sum_{i=1}^n \begin{cases}
	\ell(y_i, U_i^{\mathbf{x}}) \qquad	\text{if $y_i > U_i^{\mathbf{x}}$},\\
	\ell(y_i, L_i^{\mathbf{x}}) \qquad	\text{if $y_i < L_i^{\mathbf{x}}$},\\
	0 \qquad \qquad \quad \text{otherwise.}
\end{cases}
\]

\section{Approach}\label{ch:4-approach}

\subsection{Challenges for Authentication}

Related works demonstrate that methods are known for generating synthetic data to augment the training of activity recognition classifiers. However, authentication tasks require discriminating between similar actions by different users, and so are a more involved set of tasks that have not yet been addressed by the literature.

The cheapest and least complex approach to generating synthetic data is applying class-invariant transformations to real-world data. Unfortunately, it is not well understood which transformations preserve class labels for authentication using gesture data. Common timeseries transformations such as jittering, temporal scaling and intensity scaling have been shown not to preserve class labels in the related context of user identification~\cite{beneguiAugmentNotAugment2020}.

As a result, we considered two of the most popular deep generative frameworks~\cite{iglesiasDataAugmentationTechniques2023}, namely adversarial models such as GANs and autoencoder-based models such as VAEs. While GANs typically generate data of higher fidelity than autoencoder-based models, they frequently suffer from unstable training~\cite{goodfellowGenerativeAdversarialNetworks2020} and mode collapse~\cite{zhangModeCollapseGenerative2021a}. These issues proved insurmountable in our preliminary tests, so attention was restricted to autoencoder-based models.

\subsection{Our Approach}\label{sec:our_approach}
We are in the setting where we have a large number of gestures from existing users and a small number of gestures from a target user. Our research question is as follows: \textbf{How can we generate additional realistic gestures for the target user?}

To use an autoencoder to generate synthetic data, we must first train the encoder and decoder together to embed gestures into a latent space and reconstruct gestures from their embeddings. After training, points can be sampled from the learnt latent space and passed through the decoder to generate synthetic gestures.

To generate user-specific gestures successfully, an autoencoder model must satisfy the following two requirements:

\begin{enumerate}
	\item High reconstruction quality --- the decoder must reconstruct realistic gestures from embeddings in the latent space.
	\item Meaningful latent space representations --- the encoder must learn a latent space representation where nearby embeddings decode to semantically similar gestures (ideally gestures from the same user).
\end{enumerate}

To achieve this, we first build an autoencoder-based generative model with high reconstruction quality and meaningful latent space representations. We then use the generative model to generate synthetic gestures for a particular user and evaluate the usefulness of our method.

As our trained generative model will not have been exposed to the new user, we rely on the assumption that given a large enough corpus of users we may find a latent space that adequately captures common discriminative features between users. We select points close to a user's previously seen latent space embeddings to decode, with the intention of retaining user-specific gesture characteristics.

We call our final system for generating additional user-specific synthetic data \textit{UserBoost}.

\subsection{Threat Model}

Throughout this paper, we adopt the threat model of WatchAuth~\cite{sturgessWatchAuthUserAuthentication2022}. That is, we assume that an adversary has possession of a legitimate user's unlocked smartwatch, and that the adversary attempts to make a contactless payment using it. This is a zero-effort attack, as the adversary is attempting to fool the authentication system by submitting their own biometric signal.

\subsection{Evaluating Authentication Models}

Given biometric data, a biometric authentication system outputs a probability that the data belongs to a given user. If this probability is above a threshold $ T $, the system accepts the user; otherwise, it rejects them. Security may be balanced against usability by modifying the threshold value $ T $.

In the case of WatchAuth, maximum usability is desirable. The relevant metric in this case is the False Acceptance Rate (FAR) when $ T $ is optimised to set the false rejection rate to 0, i.e. "the false acceptance rate when we permit no false rejections". Henceforth, we abbreviate this quantity to FAR@0.

\begin{figure}[t]
	\centering
	\includegraphics[width=0.5\linewidth]{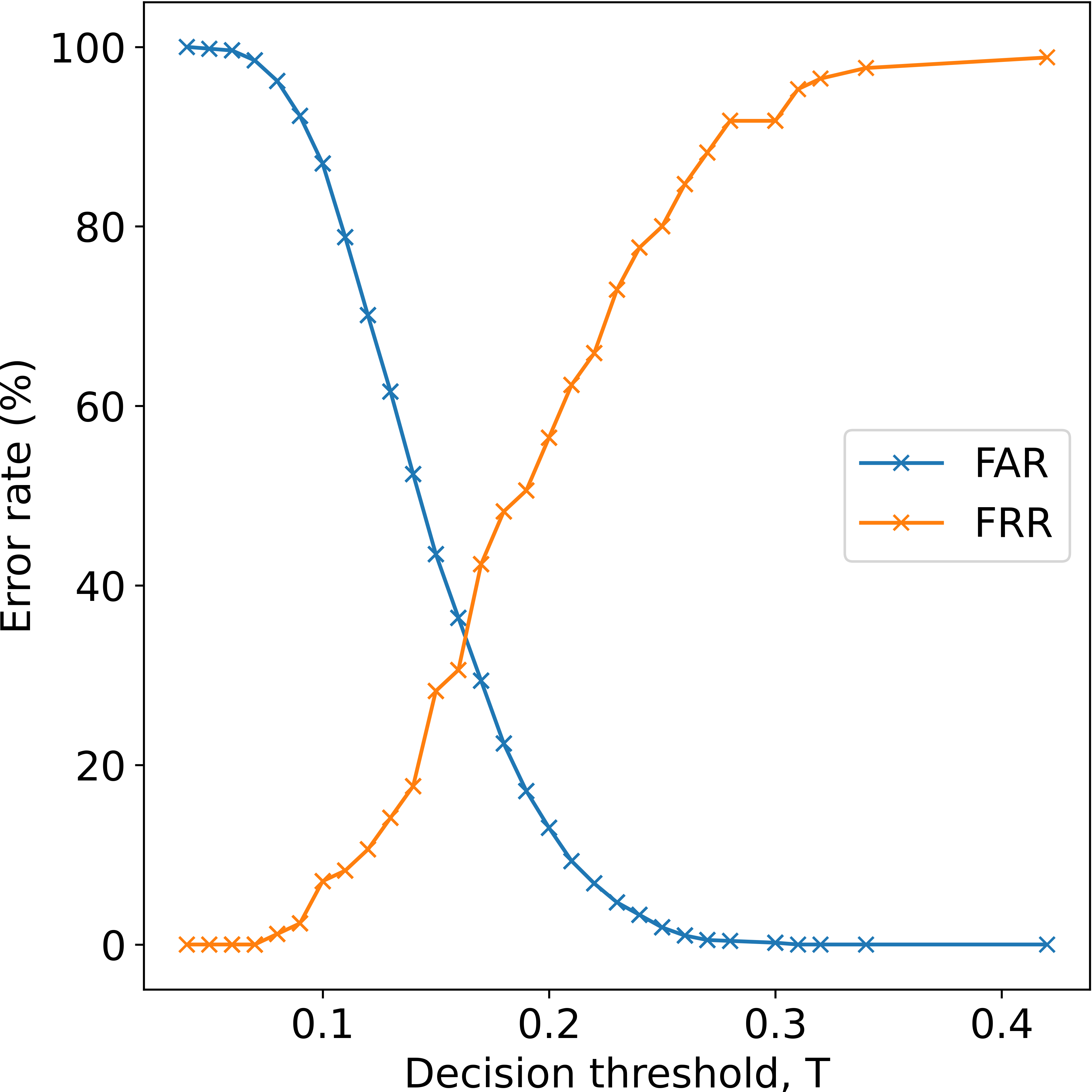}
	\caption{\small A plot of FAR and FRR against decision threshold $ T $ from a \textit{RF100} classifier (defined in Section~\ref{sec:RF100}). Marked points show thresholds. Observe that the true EER may lie between $ \sim30\% $ and $ \sim45\% $.}
	\label{fig:eerdiagramdegenerate}
\end{figure}

For models trained on limited data, many test samples may be predicted the same probability, particularly for random forest models whose implied probabilities are output from a discrete set. This poses a problem when calculating Equal Error Rate (EER), as the crossover point between FAR and False Rejection Rate (FRR) can lie in a large interval (see Figure \ref{fig:eerdiagramdegenerate}). For reporting transparency, in several tables we present the EER as an interval, with the lower value representing the minimum value of FAR where FAR is greater than FRR, and the upper value representing the FRR value at this same threshold.
We also report the Area Under Receiver Operating Characteristic curve (AUROC) value as a secondary metric for evaluating the average authentication ability of a classifier.

\subsection{Evaluating Generative Models} \label{sec:tstr}

We may evaluate the quality of an autoencoder's reconstructions via its validation reconstruction loss.
However, this metric is not guaranteed strong correspondence with the usefulness of reconstructed gestures for training an authentication system.
Instead, the quality of reconstructions was evaluated quantitatively by the \textit{train-synthetic test-real} (TSTR) method (see~\cite{yadavQualitativeQuantitativeEvaluation2023a}, or "GAN-train" in~\cite{figueiraSurveySyntheticData2022a}).
This is a strategy for evaluating the similarity between synthetic data and true data by training a classifier on the synthetic data and testing it on the real test dataset.
Strong classification performance demonstrates explicitly that the synthetic data is of sufficiently high quality to be useful as training data.


\section{Methods}\label{ch:methods}

\subsection{Data Preparation}

The WatchAuth dataset is provided as per-user files of partially-processed gesture data. Each row is associated with a gesture ID, timestamp and one of the watch's four sensors with sensor readings.

A standard input format for timeseries prediction models is a real vector of dimension $N_{timesteps} \times N_{channels} $. We therefore pre-process each row to be indexed by gesture ID and timestep rather than by gesture ID, timestep and sensor. The data provided per gesture covers a timespan from 4 seconds before NFC contact to 2 seconds after. For real-time authentication, time after the payment is accepted is irrelevant, so we slice all gestures to only cover the 4 seconds before payment is made. Non-gesture data provided in the WatchAuth dataset is also partitioned into 4 second windows.

Restricting to the sensors of interest, we represent each gesture by a separate $200 \times 6$ array (3 channels each for accelerometer and gyroscope data).
Each gesture array is then passed through a Butterworth low-pass filter to remove noise. We also store descriptive information for each gesture, including user ID and payment terminal position.

\subsection{Experiment Details}

The choice of experiment setup can strongly influence reported results. To align with best practice, we preserved the temporal order of samples when constructing train/test splits. The first two-thirds of all users' gestures was used for training and validation, with the final third used for testing. For authentication classifiers, weighting the training data 4:1 in favour of the positive class showed the strongest classification performance on validation data.

For the timeseries data used by deep learning models, channel-wise means and variances were initialised using the training data and used to normalise all data prior to classification.

All models were implemented in TensorFlow~\cite{martinabadiTensorFlowLargeScaleMachine2015}, with the exception of one model using the Soft-DTW loss function, which was implemented in PyTorch~\cite{paszkePyTorchImperativeStyle2019} (for compatibility with Maghoumi et al.'s CUDA implementation of Soft-DTW~\cite{maghoumiDeepNAGDeepNonAdversarial2021}).

All deep learning models were trained with the Adam optimiser (with learning rate $1e-4$). Hyperparameters were tuned manually to minimise average validation loss. During training, validation loss was used for model selection (using early stopping with a patience of 150 epochs).

\subsection{Authentication Models}

To perform TSTR as described, we require some independent models for performing authentication. We consider both a Random Forest model, similar to WatchAuth, and an alternative machine learning architecture with additional expressive power (similar to the encoder in our autoencoder model)\footnote{These models were chosen as they exhibited the best authentication performance on the unaltered dataset.}.

\paraHeading{RF100} \label{sec:RF100}
Sturgess et al.~\cite{sturgessWatchAuthUserAuthentication2022} showed that the accelerometer and gyroscope are the most informative sensors for authentication. Furthermore, as many smartwatches only have these two sensors, this study restricts the data considered to just the acceleration and angular velocity vectors. In the rest of the paper, we refer to the WatchAuth random forest authentication system with 100 trees using just accelerometer and gyroscope data as \textit{RF100}. (We also investigated the equivalent system with 1000 trees, but observed only minor gains that did not justify the additional resources required for training and classification with forests of this size.)

\paraHeading{Conv+GRU}\label{sec:Conv+GRU}
We wish to construct a deep learning architecture that exploits both the timeseries nature of our input data, as well as the relationship between different channels.

Empirically, we obtained the best authentication performance by combining convolutional and Gated Recurrent Unit (GRU) layers in sequence.
Intuitively, this architecture can first extract translation-invariant features using convolutions, and then exploit intra-sequence dependencies with the GRU\footnote{GRUs were found to greatly outperform the more complex LSTM (Long Short-Term Memory) units, which could not be trained effectively.}. This approach also mitigates training difficulties when using GRUs for long sequences as the convolutional layers reduce the spatial dimension of the data.

Our \textit{Conv+GRU} model employs varying kernel sizes to extract features at multiple scales (similar to Chrononet~\cite{royChronoNetDeepRecurrent2019a}). These features are then concatenated and passed through 1x1 kernels to reduce the overall number of parameters. Four layers of this architecture are employed to reduce the spatial dimension of the data to thirteen timesteps. Data is then passed through three stacked GRUs before a classification multi-layer perceptron with two hidden layers, one with 25 units and one with 10 units, both using the Rectified Linear Unit (ReLU) activation function. An output neuron using sigmoid activation then is used to output a classification probability.

\subsection{Autoencoder Architecture} \label{sec:gen_architecture}

We now seek to construct an encoder-decoder pair for modelling payment gestures.

We take our encoder to be the \textit{Conv+GRU} model described in Section \ref{sec:Conv+GRU}. However, we remove the output sigmoid neuron so that its output is a vector of dimension equal to the latent space size (the latent embedding vector).

We construct our decoder as an inverted version of the encoder. Our decoder consists of stacked GRU layers acting on the latent embedding vector, followed by a sequence of upsampling and convolutional layers to reconstruct a full-length output signal.

The size of an autoencoder's latent space must be a compromise between the two requirements of Section \ref{sec:our_approach}, with larger spaces favouring reconstruction quality and smaller spaces favouring more meaningful latent space representations. Following extensive experimentation, the latent space size was fixed at 10 dimensions.

\subsection{Autoencoder Reconstruction Loss}

For synthetic data to be useful, it must be of sufficiently high quality to improve a classifier's performance on an authentication task. For an autoencoder, the quality of gestures is determined by the fidelity of the decoder's outputs. Given that the decoder is trained to minimise reconstruction loss, the choice of loss function strongly influences the quality of reconstructed gestures.

In the absence of a literature consensus on the choice of loss function for timeseries comparison, we investigate the effectiveness of several choices:

\begin{itemize}

\item{\textit{Mean Squared Error (MSE):}} A naive choice of loss is pointwise MSE. MSE only captures the similarity between two signals in the intensity dimension.

\item{\textit{Soft-DTW:}} DTW is an alternative quantity for measuring timeseries dissimilarity that is invariant to translation and warping in the time dimension.
However, DTW is known to perform poorly as a loss function because the \textit{min} operation used to calculate it is not differentiable.
In 2017, Cuturi \& Blondel introduced Soft-DTW~\cite{cuturiSoftDTWDifferentiableLoss2017} as a smoothed version of DTW that replaces \textit{min} with a differentiable \textit{soft-min} operation.
The level of smoothing of the \textit{soft-min} operation is controlled by a hyperparameter $ \gamma $, which we set to a standard value of $0.1$.
We calculate Soft-DTW separately for each channel --- otherwise, the time warping heavily penalises reconstructions that slightly misalign peaks for two separate channels.
We use Maghoumi et al.'s open-source CUDA implementation for PyTorch~\cite{maghoumiDeepNAGDeepNonAdversarial2021}, which follows the formulation of Soft-DTW given by Zhu et al.~\cite{zhuDevelopingPatternDiscovery2018}.

\item{\textit{Keogh's lower bound for DTW:}}
Keogh's lower bound for DTW is much more efficient to compute than Soft-DTW.
Furthermore, it is tolerant of small translational shifts / warping within its Sakoe-Chiba bandwidth.
However, it provides zero reconstruction loss when one curve is entirely within the envelope of another curve, which would limit an autoencoder's reconstruction resolution to the coarseness of the bandwidth.
We overcome this issue through a modification of Keogh's lower bound, which we name \textit{KLB-mod} and construct as the weighted average of Keogh's lower bound at exponentially varying bandwidths:
\[ \qquad d_{\text{\textit{KLB-mod}}}(x,y) = \sum_{w=1}^{5} (6-w) \cdot d_{KLB}^{(2^w)}(x,y) \]
where $ d_{KLB}^{(w)}$ is Keogh's lower bound.
This formulation results in a theoretically smoother loss landscape than MSE or Keogh's lower bound with fixed bandwidth.

\item{\textit{Feature-based approaches:}}
WatchAuth compares gestures based on their feature similarity rather than on explicit differences between the original curve amplitudes.
To reconstruct gestures that are highly discriminative for authentication, we consider augmenting the above reconstruction losses with a feature-based loss.
We represent a curve as a vector of differentiable features and apply a standard Euclidean kernel.
The features used in WatchAuth that are compatible with TensorFlow are used for this feature vector representation, namely \textit{max}, \textit{min}, \textit{mean}, \textit{standard deviation}, \textit{variance}, \textit{skew}, \textit{kurtosis}, \textit{median} and \textit{inter-quartile range}.

\end{itemize}

We trained autoencoders with each of the reconstruction losses above, as well as the combination of feature-based loss and each other loss. We then generated synthetic data with the trained autoencoders, and evaluated this synthetic data with a TSTR task classifying timeseries data into gesture and non-gesture classes (details in Appendix \ref{app:losses}).
\textit{KLB-mod + Feature}, a loss additively combining \textit{KLB-mod} and our \textit{Feature} loss achieved the best empirical performance and so we use this loss function going forwards.

\begin{figure}[t]
	\centering
	\includegraphics[width=0.9\linewidth]{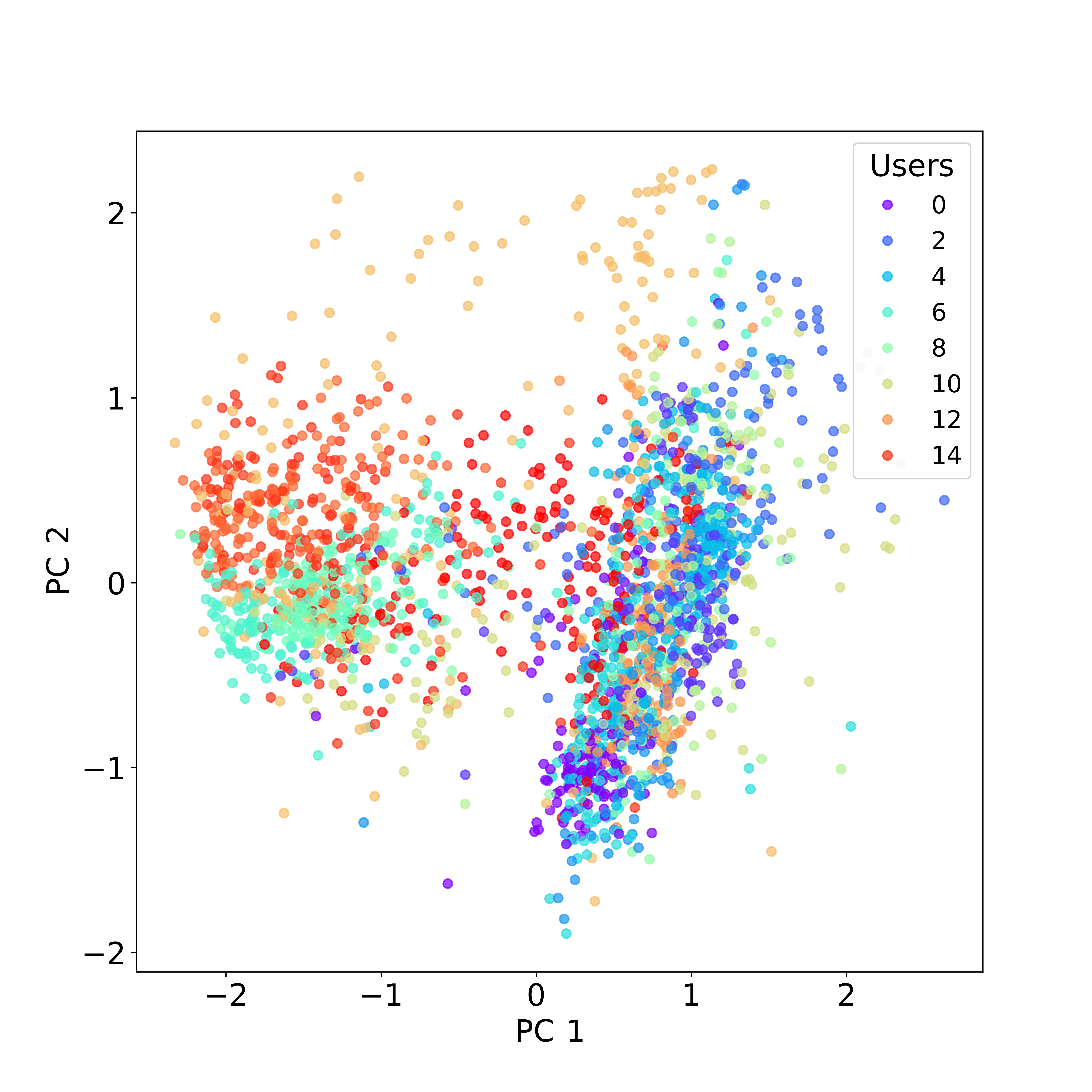}
    \vspace{-5mm}
	\caption{\small The first two principal components (PCs) of the training data's latent space embedding for an unregularised autoencoder trained using the \textit{KLB-mod + Feature} loss.}
	\label{fig:latentspaceunregularised}
\end{figure}

\begin{figure}[t]
    \centering
    \includegraphics[width=1\linewidth]{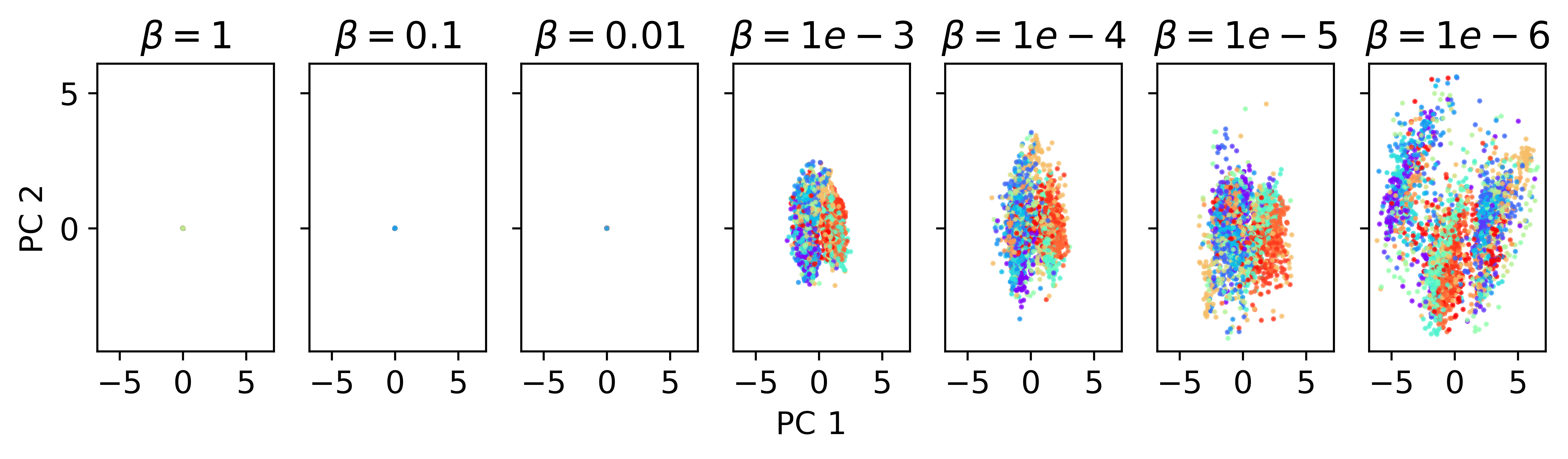}
    \caption{\small Training data latent space mean embeddings for VAEs trained with decreasing $ \beta $ values, decreasing the influence of regularisation. Different colours represent different users.}
	\label{fig:beta_latent_spaces}
\end{figure}

\subsection{Regularising the Autoencoder}\label{sec:autoencoder_regularisation}

Figure~\ref{fig:latentspaceunregularised} visualises the latent space of an autoencoder model trained using \textit{KLB-mod + Feature}. It is promising that there appears to be a degree of natural clustering by user. However, the latent space lacks the necessary continuity to be useful for generating synthetic gestures and so requires regularisation.


We used the VAE framework for regularisation, which introduces a KL regularisation term into the autoencoder loss function, weighted by scalar hyperparameter $\beta$:
\[ \qquad \ell_{VAE} = \ell_{reconstruction} + \beta \cdot \ell_{regularisation} \]

We exponentially varied the value of $\beta$ between 1 and $ 1e-6 $ and trained a VAE on the training dataset for each $ \beta $.

Figure~\ref{fig:beta_latent_spaces} visualises how the latent space becomes less regularised as $ \beta $ decreases. We can see that, for $ \beta \ge 1e-2 $, all points are encoded with approximate mean 0 and standard deviation 1. For $ \beta \le 1e-3$ regularisation still has an effect but is balanced with reconstruction ability (see Appendix \ref{app:WAE}, Figure \ref{fig:waesvallosses}).
As a result we set $ \beta = 1e-4 $.

Appendix \ref{app:WAE} details an empirical comparison of Wasserstein autoencoders to VAEs as a regularisation strategy.

\subsection{Enforcing User Similarity in the Latent Space}

To be able to sample points from the latent space and have confidence in their user label, we must achieve much more consistent clustering by user in the latent space than has been observed in the models so far. This motivates the explicit enforcement of user-based clustering in the latent space.

To this end, we add a simple Multi-Layer Perceptron (MLP) classifier to our VAE model, and train it to classify points in the latent space according to their user. We strictly limit the expressive power of this classifier. As a result, to improve the latent space user classification the encoder must learn to organise the latent space by user.
Under the hypothesis that the distribution of gestures is determined both by user and non-user variables (e.g. terminal position, independent random effects), we only apply this classifier to the first 5 dimensions of the 10-dimensional latent space.

\begin{figure}[t]
	\centering
    \hspace{-1.0cm}
	\includegraphics[width=1.1\linewidth]{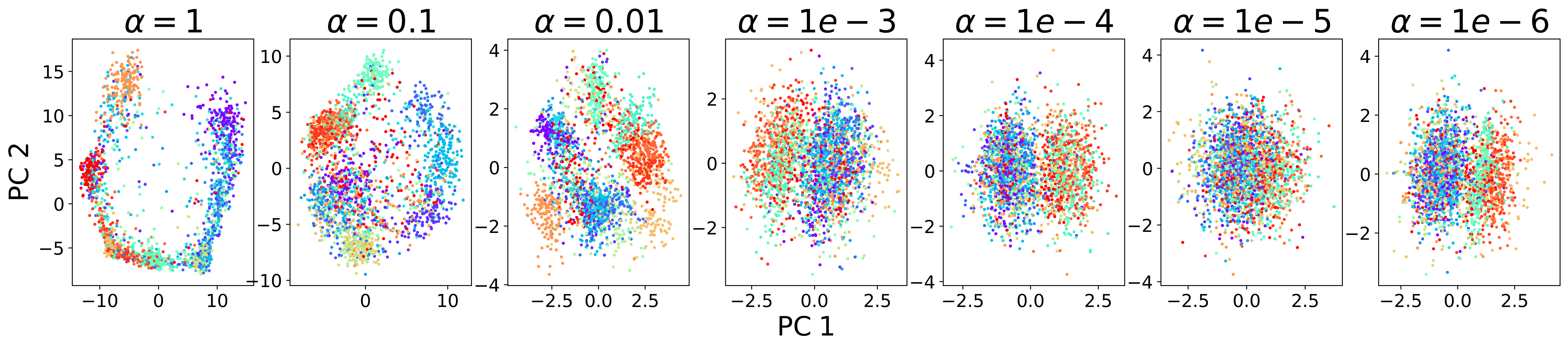}
	\caption{\small Latent space visualisation for VAE-based generative models trained with varying values of $\alpha$, increasing the emphasis put on a structured latent space in the modified VAE loss function.}
	\label{fig:alphazs}
\end{figure}

The natural loss function for this classifier is categorical cross-entropy. Empirically, we observed that this loss caused the model to increase its confidence in points it was correctly classifying rather than focusing on classifying all points correctly. An alternative choice of loss that we found to induce stronger clustering was a rank-based loss function, where points in the latent space are assigned scores per user, and the objective is to maximise the Mean Reciprocal Rank (MRR) of the true user (to reward higher predicted ranks to the true user).
For differentiability and compatibility with our deep learning architecture, we used \textit{approximate} MRR~\cite{qinGeneralApproximationFramework2010} and implemented in TensorFlow Ranking~\cite{pasumarthiTFRankingScalableTensorFlow2019}. 
This latent space authentication loss leads to a new loss function for our model:
\[ \qquad \ell_{total} = \ell_{\text{\textit{KLB-mod + Feature}}} + \beta \cdot \ell_{KL} + \alpha \cdot \ell_{auth}\]

We introduce a new hyperparameter $ \alpha $ to control the relative effect of the authentication loss. As with $\beta$ in Section~\ref{sec:autoencoder_regularisation}, we experimented with different values of $ \alpha $ to balance reconstruction quality, regularisation and classification by user. Figure~\ref{fig:alphazs} shows the latent space at varying values of $ \alpha $. Note that setting $ \alpha $ too high results in excellent grouping by user and very poor regularity.

We set $ \alpha = 1e-2 $, guided by the effect on reconstruction loss (see Appendix \ref{app:auth_loss},  Figure~\ref{fig:valreconlossauthvaes}). The validation approximate MRR is $ \approx 0.1$ without and $ \approx 0.3$ with the authentication MLP, which shows it induces an improvement in clustering by user.

\subsection{Model Design Summary} \label{sec:vae_design}

Informed by the preliminary experiments and discussion in this section, we constructed the following autoencoder model to generate synthetic payment gestures. We constructed a VAE-based model, with encoder and decoder as specified in Section~\ref{sec:gen_architecture}, with the following weighted sum loss:

\begin{itemize}
    \item Reconstruction loss ($\times 1$): the sum of a feature-based loss, to reward reconstructions that preserve key features of curves such as $\max$ and $\min$, and the weighted average of Keogh's lower bound at exponentially varying bandwidths, to smoothly reward faithful reconstructions.
    \item Regularisation loss ($\times 1e-4$): KL-divergence (as normal for VAE), acting on magnitude of latent neurons in the model.
    \item Latent space authentication loss ($\times 1e-2$): a rank-based loss penalising latent spaces that do not display sufficient organisation by user. A simple MLP is trained in parallel with with main network to classify users from their latent space embeddings (only using half of the latent space neurons), and the loss is applied to these predictions.
\end{itemize}

\section{Results}\label{ch:experiments}

\subsection{Verifying Model Properties}

\begin{figure}[t]
	\centering
	\includegraphics[width=0.8\linewidth]{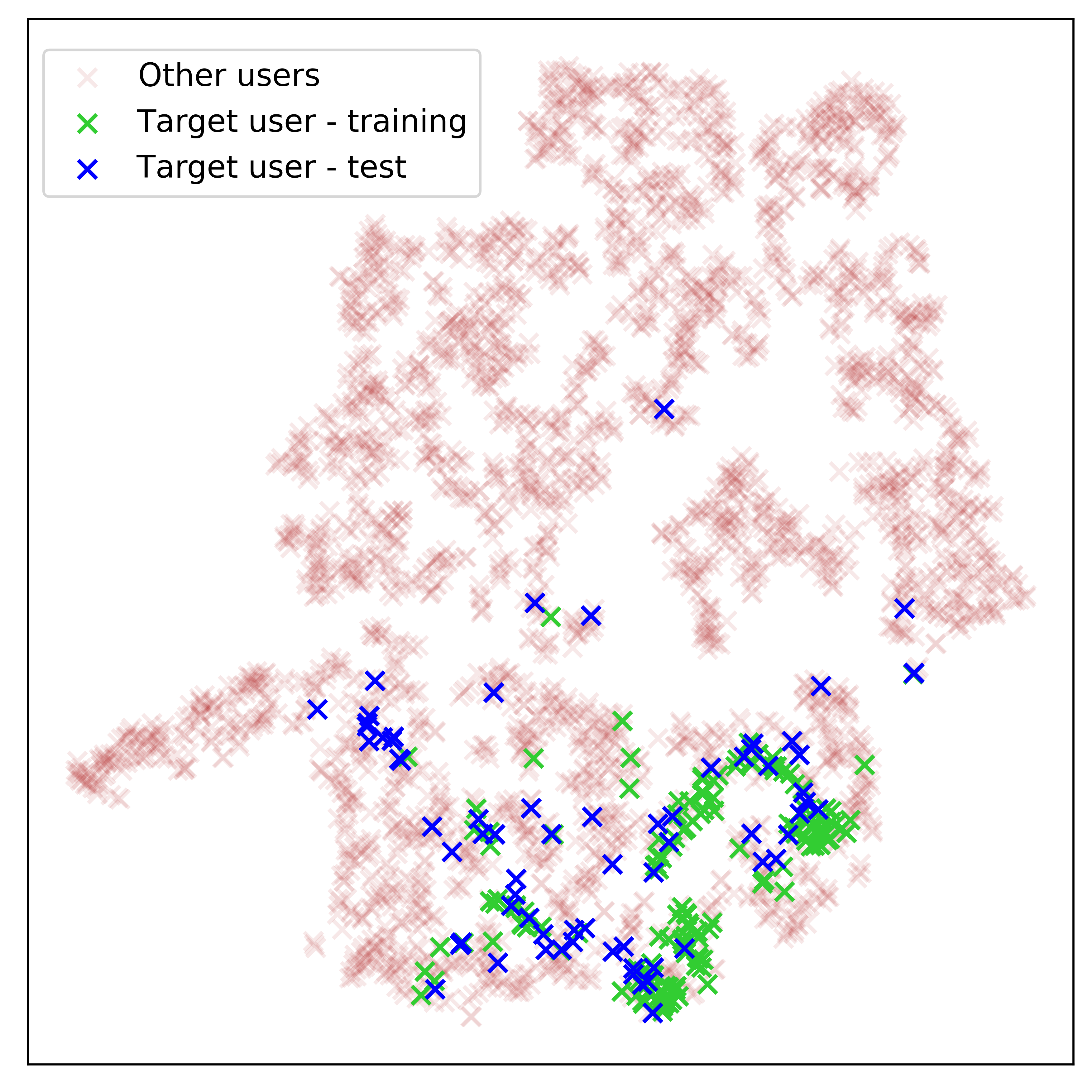}
	\caption{\small t-distributed Stochastic Neighbour Embedding (t-SNE) visualisation of a trained VAE-based model's latent space, demonstrating that an example user's test gestures are embedded similarly to the user's training gestures.}
	\label{fig:unseenembeddings}
\end{figure}

\paraHeading{Latent space structure}
For synthetic gestures to be useful, our encoder must assign similar embeddings to training gestures and unseen gestures for a given user. Figure~\ref{fig:unseenembeddings} shows the result of embedding one user's gestures in the test dataset for a model trained using gestures in the training dataset.

It may be observed that the gestures observe a degree of localisation to a region of the embedding manifold. It is unrealistic to expect perfect clustering given that authentication classifiers trained on this dataset do not achieve perfect classification. With more users, the quality of gesture clustering would be expected to improve.

\begin{figure}[t]
	\centering
	\includegraphics[width=0.9\linewidth]{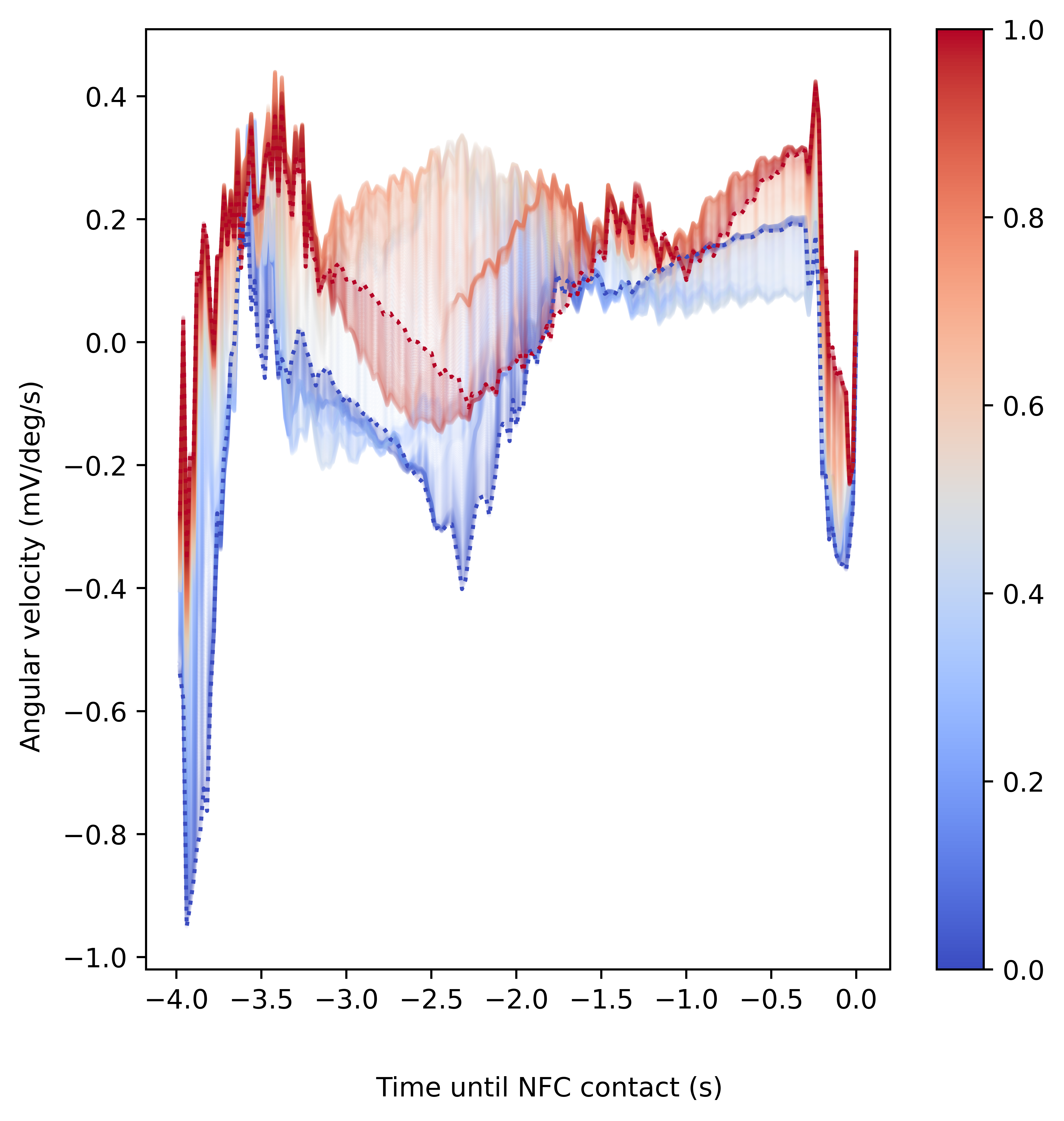}
	\caption{\small Visualisation of decoding points along a short line segment in the latent space. For points $ x $ and $ y $, we visualise the decoded gesture from latent space point $ \mu \cdot x + (1-\mu) \cdot y $ for each $ 0 \le \mu \le 1$. The reconstructions with $ \mu = 0 $ and $ \mu = 1 $ are represented by blue and red dotted lines respectively. Colours associated with other values of $ \mu $ are shown by the colour bar.}
	\label{fig:notlinear}
\end{figure}

\paraHeading{Diversity and fidelity of synthetic gestures}
We verify that the decoder is generating diverse, high fidelity samples and is not just linearly interpolating between existing gestures. This can be seen in Figure~\ref{fig:notlinear}, which we created by sampling two nearby points from the latent space, calculating points on the line between them, and decoding these intermediate points.

\subsection{Generating User-specific Synthetic Gestures}

We now seek to use our final VAE-based model (from Section~\ref{sec:vae_design}) to generate user-specific synthetic gestures, and evaluate whether these gestures are useful for an authentication use case.

Given a VAE-based model trained on other users, we can apply the encoder to embed the target user's gestures into the model's latent space. We can then use these embeddings to guide the selection of further points in the latent space, which we then decode to generate our user-specific synthetic data. 

In this section, we firstly investigate the effectiveness of four simple strategies for selecting points in the latent space. We then proceed with the most promising sampling strategy to answer the central question of whether our methodology can improve the classification performance of a simple authentication model.

\subsection{Latent Space Sampling Strategies}

\paraHeading{Neighbourhood sampling}
We simply sample points from a Gaussian distribution using the embedding means and variances for the points we already know for the user.

\paraHeading{Self-mixed sampling}
We pick several of the existing embedding means for the user, and generate a new point in the latent space by calculating a randomly weighted convex combination of these points.

\paraHeading{Adversarial sampling}
We randomly pick two embeddings: one from our target user and one from a different user. We then calculate a convex combination of them weighted heavily towards the target user (85:15). The rationale behind this strategy is to generate gestures that should be classed as the target user but have some characteristics of other users, to generate a more diverse training set.

\paraHeading{Same-user sampling}
We take existing embeddings for the target user. We assume that as the authentication MLP is trained on the first 5 dimensions of the latent space that these dimensions are user-specific and other dimensions are not. We therefore set the embedding's last 5 dimensions to randomly sampled embeddings for other users' points.


\subsection{Experiments}

To evaluate whether our methodology improves the classification performance of a simple authentication model, we performed TSTR for authentication using the classifier \textit{RF100} (defined in Section~\ref{sec:RF100}). We trained 16 models, leaving out training data for one user each time. This mimics the realistic scenario of pre-training a generative model before seeking to enrol a new user into an authentication system with the aid of synthetic gestures.

For each trained model, we allowed access to two real user gestures per terminal position (14 in total), which were used to generate 500 synthetic gestures with each of our four sampling strategies. The original and synthetic gestures together formed the positive class for training an authentication classifier. For the results in Section~\ref{sec:results_overall}, our negative class contained the VAE-based model's reconstruction of each of the other users' gestures.
We then converted our training data (using both real and synthetic gestures) into feature vectors and trained a random forest classifier for authentication.

In Section~\ref{sec:results_reduceburden}, we also included the original gestures from other users as negative samples, and varied the number of real user gesture samples per terminal to investigate how effective our approach is at reducing enrolment burden.

As before, the full test dataset of gestures unseen during training was used to evaluate all trained authentication classifiers.

\begin{table}[t]
	\caption[]{\small TSTR authentication results for a random forest classifier aided by synthetic data generated by different strategies.} 
	\label{tbl:TSTR_auth}
	\vskip 0.15in
	\begin{center}
		\begin{sc}
			\input{tables/TSTR_auth}
		\end{sc}
		\vskip -0.4in
	\end{center}
\end{table}

\subsection{Results}\label{sec:results_overall}

\paraHeading{Overall performance} 
TSTR results for authentication, using synthetic data generated with different sampling strategies, are presented in Table~\ref{tbl:TSTR_auth}.

The synthetic data provided a significant improvement to FAR@0, irrespective of sampling strategy. 
Adversarial sampling yielded the strongest performance on all metrics amongst the sampling strategies, with an average reduction in FAR@0 from 100\% to 74\% and favourable AUROC compared to using no synthetic data.
The addition of the synthetic data also reduced the uncertainty around the true EER value.
To investigate the cause of this improvement it is necessary to consider performance on a per-user basis.

\begin{figure*}[t]
    \centering
    \begin{minipage}{0.35\textwidth}
        \centering
        \includegraphics[width=\linewidth]{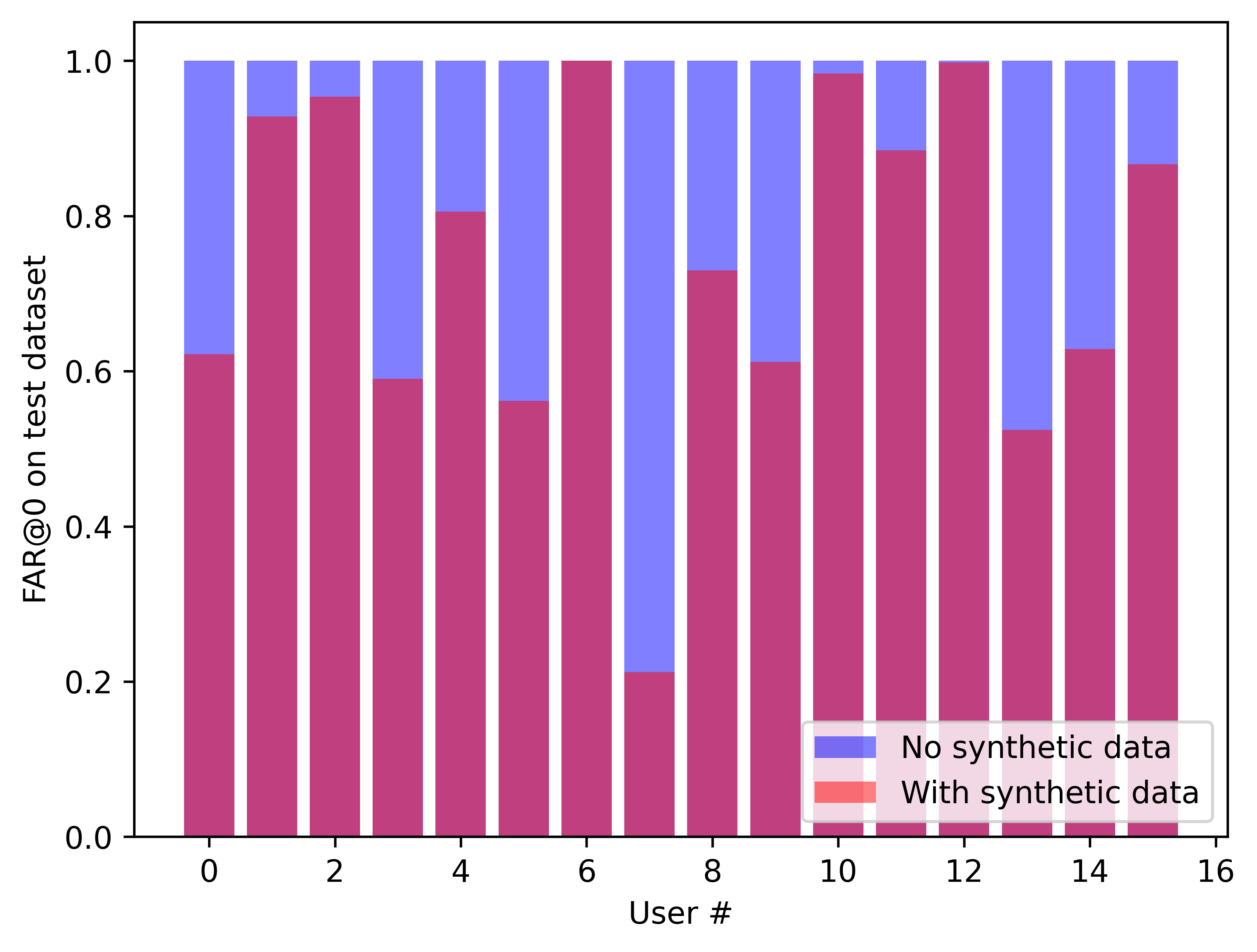}
        \caption{\small Average FAR@0 for random forest classifiers, trained with and without synthetic gestures (generated using adversarial sampling). Results are broken down by user.}\label{fig:farforadvsampling}
    \end{minipage}\hfill
    \begin{minipage}{0.6\textwidth}
        \centering
        \includegraphics[width=0.9\linewidth]{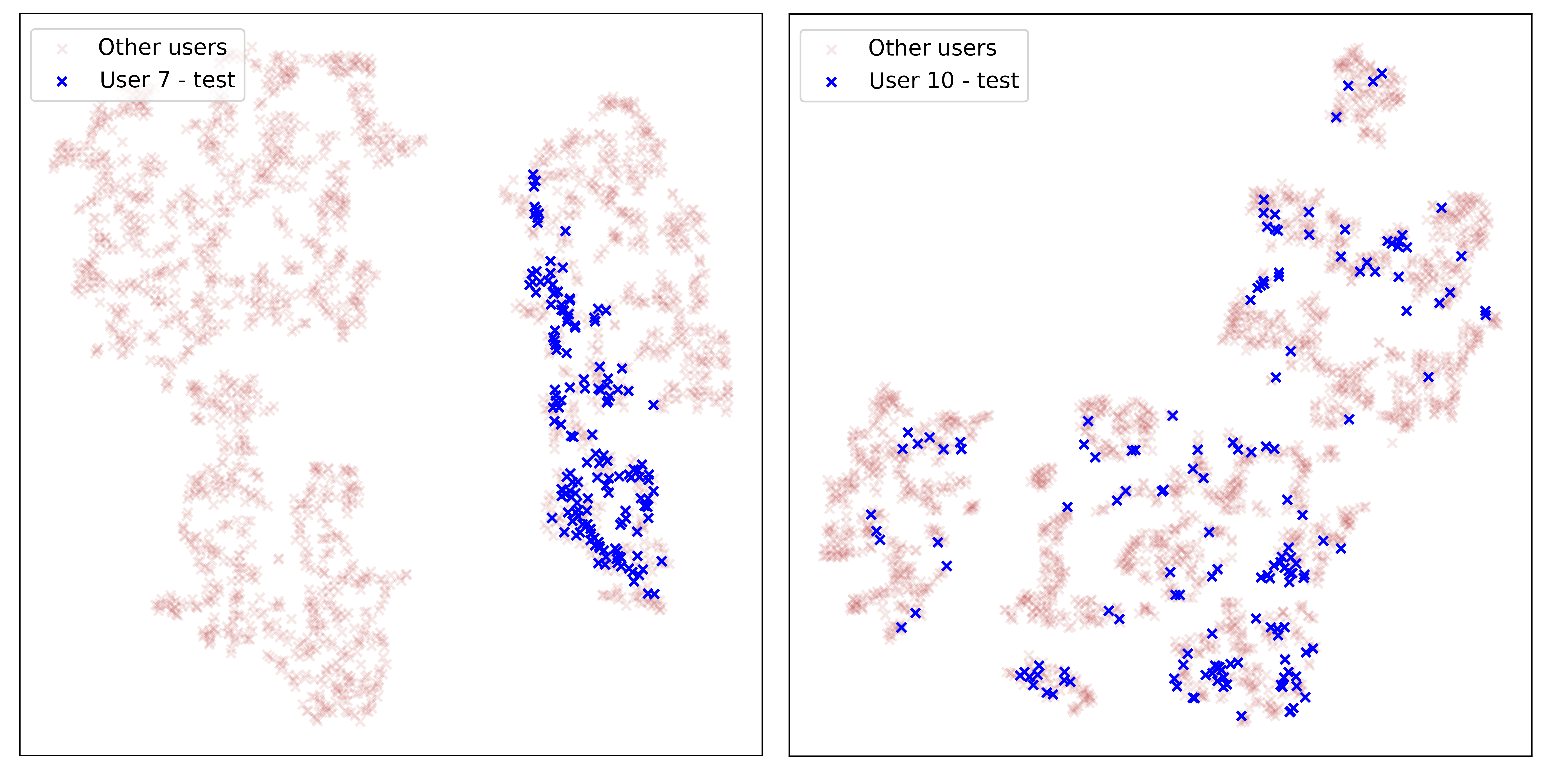}
        \caption{\small Left: User 7. Right: User 10. t-SNE visualisation of latent space embeddings of test data for User $ x $ and training data for other users, for a VAE-based model trained without User $ x $ ($ x = 7 $ or $ x = 10$).}\label{fig:usertsne}
    \end{minipage}
\end{figure*}

\paraHeading{Per-user performance}
Figure \ref{fig:farforadvsampling} shows the FAR@0 for each user when synthetic gestures were generated using adversarial sampling. It is clear that performance varied widely between users. For example, User 7 experienced a reduction in FAR@0 from $ 1.0 $ to $\sim 0.2 $, while Users 6, 10 and 12 had very little or no reduction.

We visualise the embeddings of test gestures for Users 7 and 10 in Figure \ref{fig:usertsne}. In the case of User 7, our VAE-based model trained without User 7's gestures learnt a user clustering that is highly relevant to User 7. For User 10, no such effective organisation of the latent space took place. This example shows the importance of the assumption that closely localised points are more likely to share a user. It is a positive finding that when this assumption is met, synthetic gesture data is of high quality and improves classification performance.

\subsection{Impact on Enrolment Burden}
\label{sec:results_reduceburden}

\begin{figure*}[t] 
	\centering
	\begin{subfigure}[b]{0.33\linewidth}
		\centering
		\includegraphics[width=\linewidth]{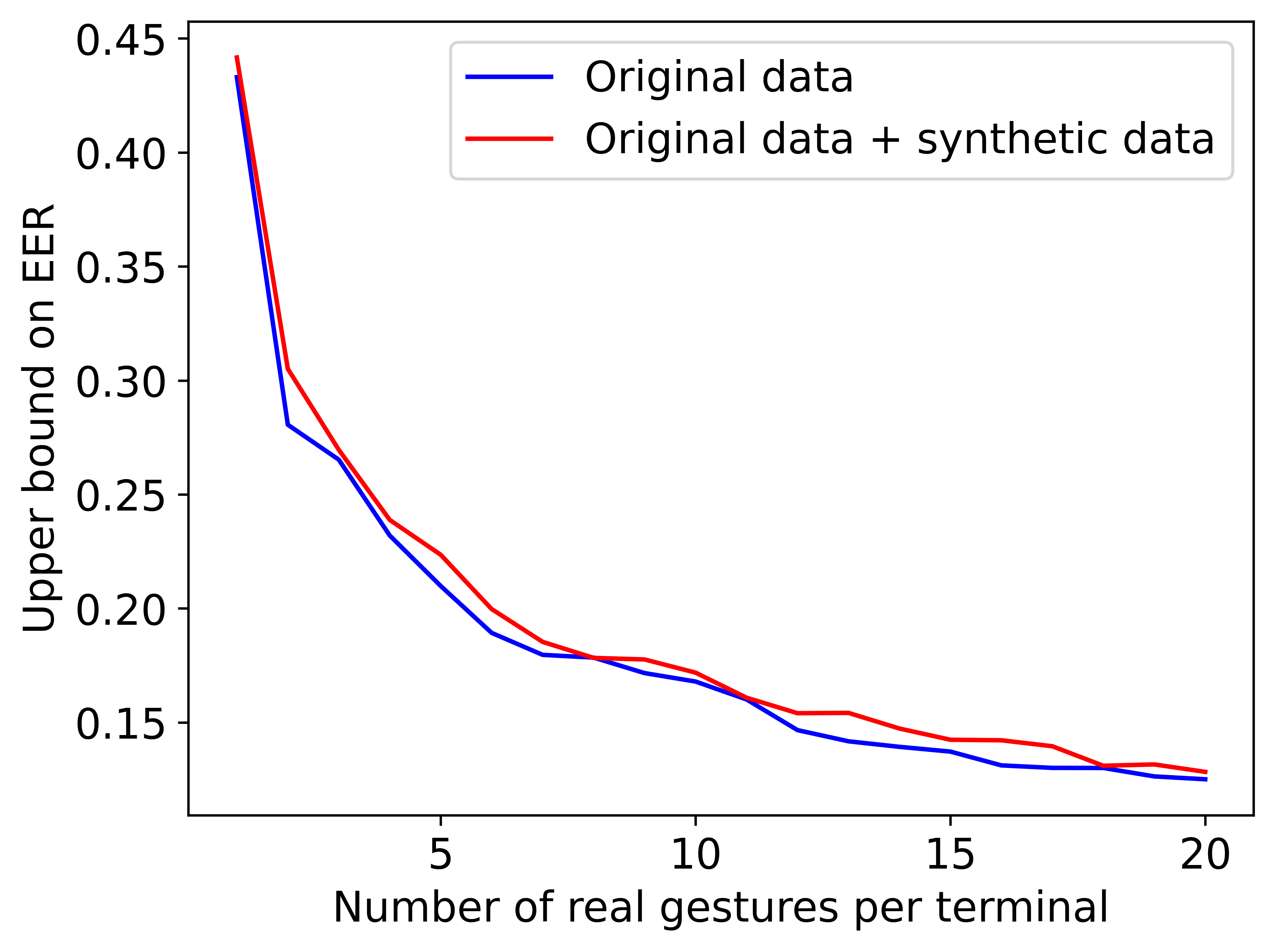}
		\caption[]%
		{{\small Upper bound on EER}}
	\end{subfigure}
        \hfill
	\begin{subfigure}[b]{0.33\linewidth}
		\centering 
		\includegraphics[width=\linewidth]{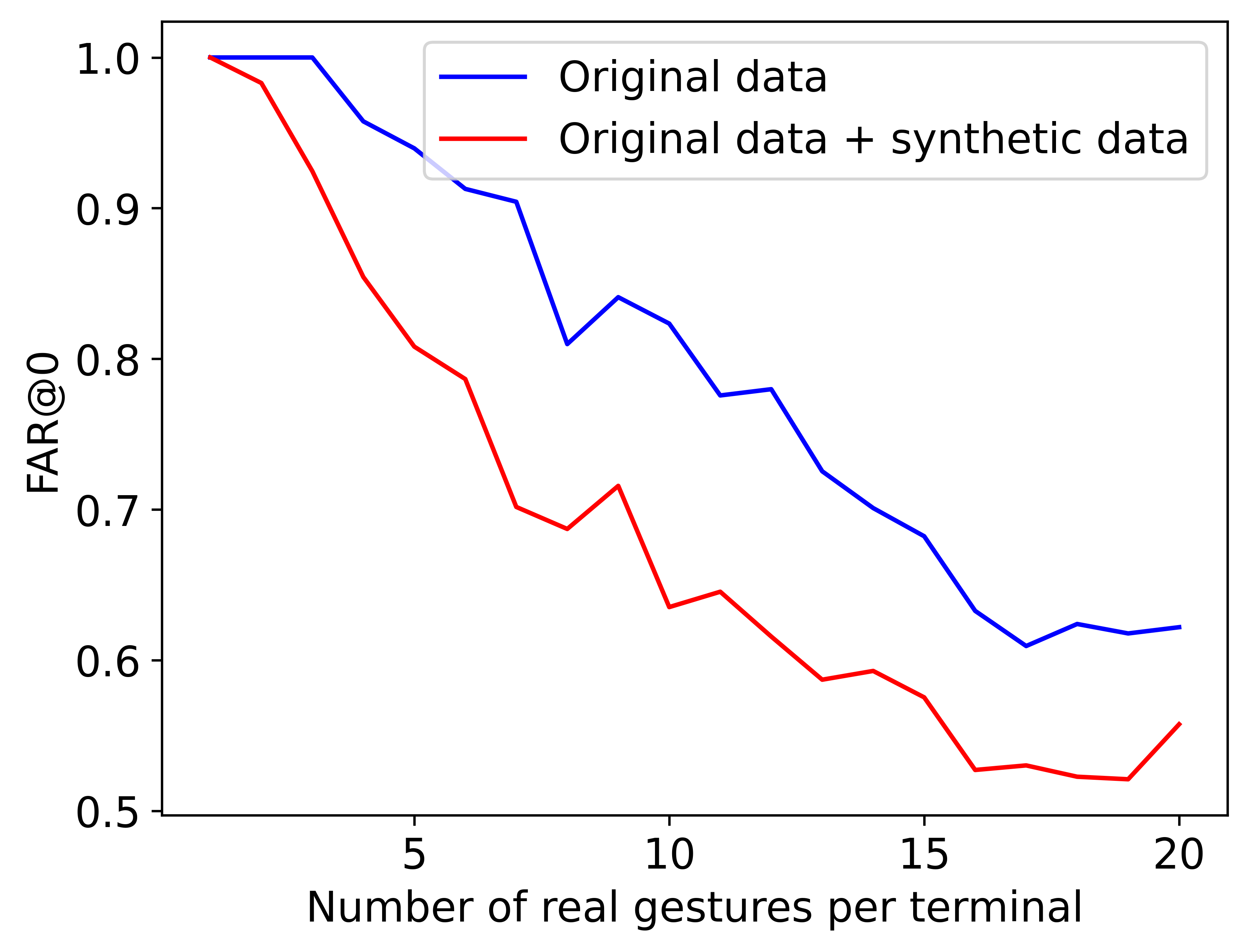}
		\caption[]%
		{{\small FAR@0}}
	\end{subfigure}
	\hfill
	\begin{subfigure}[b]{0.33\linewidth}
		\centering 
		\includegraphics[width=\linewidth]{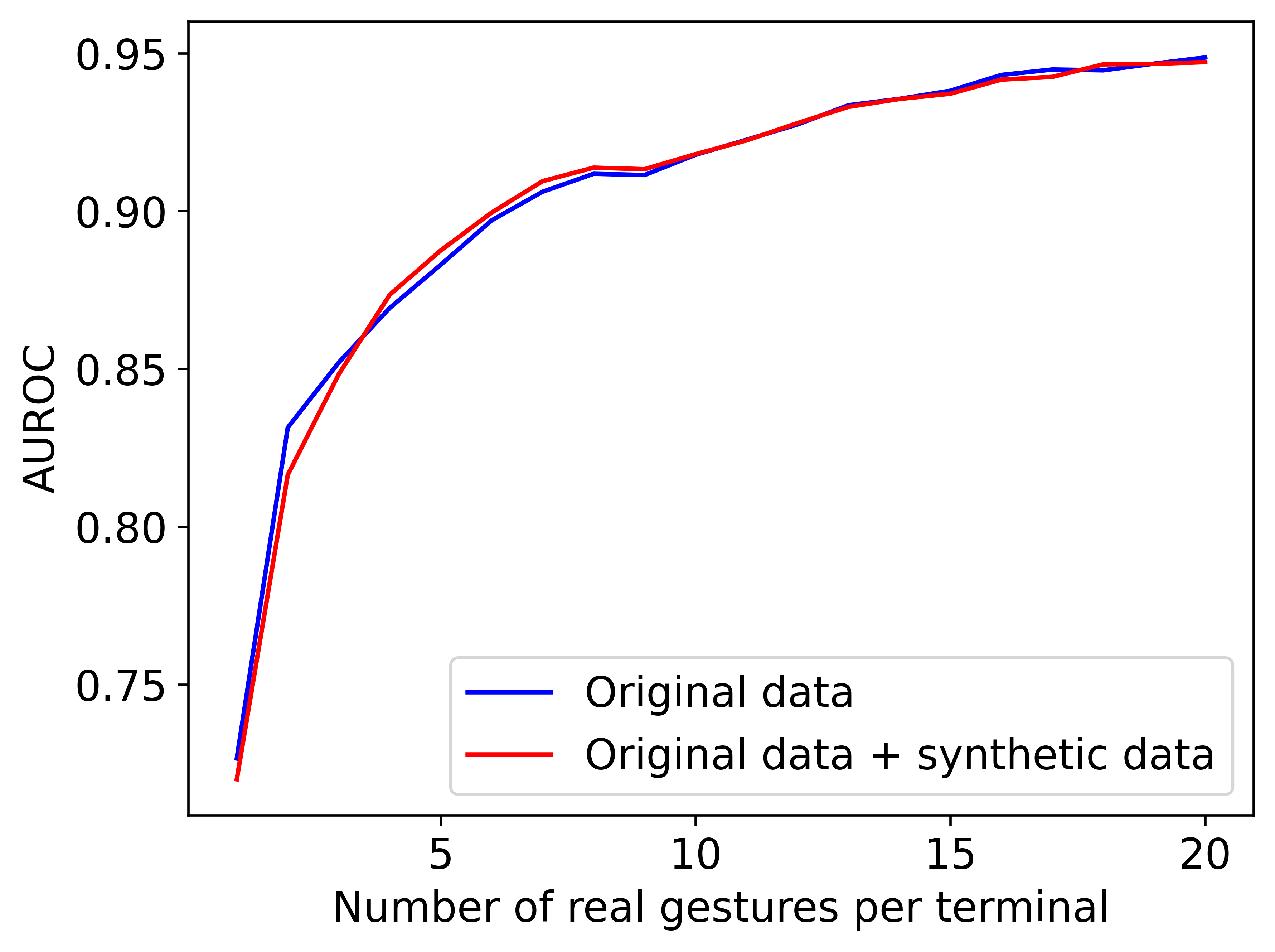}
		\caption[]%
		{{\small AUROC}}
	\end{subfigure}
	\hfill
	\caption[]%
	{{\small Authentication metrics for random forest classifiers given increasing quantities of training data, with and without the addition of synthetic data.}}
	\label{fig:reduceburden}
\end{figure*}

The primary aim of this work was to explore whether using synthetic gestures can reduce the enrolment burden of WatchAuth. To this end, in Figure~\ref{fig:reduceburden} we visualise how increasing the number of real user gestures used in training affects the quality of an authentication classifier.
We observe that the addition of synthetic gestures does not adversely affect the EER and AUROC of our trained classifier. Furthermore, the FAR@0 is markedly improved.

It is clear from Figure~\ref{fig:reduceburden} that the enrolment burden on a user can be reduced with our approach. For example, with synthetic gestures, only 9 gestures per terminal were required to reach a FAR@0 value of 70\%, while without synthetic gestures 16 gestures per terminal were needed, an improvement of 42 fewer gestures (in total) and a reduction of over 40\%.

\section{Discussion and Limitations}\label{ch:7-discussion}

\paraHeading{Variation in performance between users}
In Section~\ref{sec:results_overall} we showed a large disparity between the effectiveness of our methodology for enrolling different users. This could be due to fundamental differences between the distinctiveness and stability of user gestures. However, it is more likely to be because some users are more similar to others in the dataset collected, or a result of improper data collection for some users. Both when training simple authentication models on real data and when incorporating synthetic data, we observed large differences in the quality of authentication for different users, with a small subset of users proving difficult to authenticate.

If a user provided a poor quality gestures, the synthetic data for that user would also likely be poor quality; however, as more real gestures were provided, the authentication model would be retrained so this does not detract from the benefits we show.

\paraHeading{Improvement of average measures}
FAR@0 is a measure specific to our use case of minimising enrolment burden, prioritising security over usability. The more general measures of EER and AUROC capture a blend of the authentication system's security and usability. Our results did not show any improvement in EER or AUROC when adding synthetic data to the training process for \textit{RF100}.

The sampling methodologies used are inherently likely to improve usability at a potential cost to security, because there are no guarantees over what points in the model's latent space are associated with which users. As a result, it is a likely scenario that models are made more permissive by the inclusion of training data that shares characteristics associated with other users.

However, with a larger corpus of users there remains the possibility of improving security as well as usability through the exploitation of commonalities between user groups.


\paraHeading{Latent space assumptions}
The success of our synthetic data generation methodology rests strongly on the assumption that we can find a latent space of gesture data embeddings in which the gestures of unseen users cluster meaningfully. This work has not fully validated that assumption, although the success of synthetic gesture generation for some users is promising (especially when considering Figures \ref{fig:farforadvsampling} and \ref{fig:usertsne} together).

\paraHeading{Synthetic gesture generalisability}
Qualitatively, we found that our reconstructed gestures were sometimes of lower fidelity than their corresponding real gestures. As the gestures reconstructed by our generative model were only tested for our authentication enrolment use case it remains unknown how useful they are in other contexts. Additional evaluation would clarify the wider usefulness of our synthetic gestures, while integrating this work with GANs or diffusion models provides an avenue to generating higher fidelity synthetic data.

\paraHeading{Other use cases}
The methodology employed in this work could be applicable to other hardware-constrained authentication tasks. It also has potential as a student-teacher learning framework for reducing the memory requirements of authentication models.

\section{Conclusion}\label{ch:8-conclusion}

We have shown the feasibility of generating synthetic IMU data from scratch corresponding to a specific user's physical gesture using a deep-learning pipeline.
When used to augment the training process of WatchAuth, a simple machine-learned biometric authentication system, this synthetic data reduces the enrolment burden for a user by up to $ 40 \% $ without compromising security. It also reduced the false acceptance rate when no false rejections are permitted by an average of $ 26 \% $ for a case of minimal training data. Once trained, our synthetic data generation solution is efficient and generates diverse synthetic data.

The usefulness of the synthetic data varied between users, with some users seeing excellent results with synthetic data and others only low to moderate improvements.
It is hoped that larger studies would be better able to exploit the potential of transferring learnt semantic information from a large corpus of users' data to the authentication template for a new user.

\section*{Acknowledgements}

This work was supported financially by Mastercard and the\\Engineering and Physical Sciences Research Council [grant number EP/P00881X/1]. The authors would like to thank these organisations for their support.

\bibliography{bibtex_refs}

\section*{Appendix}

\appendix

\section{Autoencoder Reconstruction Loss Experiments}\label{app:losses}

\paraHeading{Preliminary experiments}
We conducted experiments with loss functions defined as the weighted sum of this feature loss and each of the MSE and KLB-mod loss functions, which we call \textit{MSE + Feature} and \textit{KLB-mod + Feature} respectively.
The relative weight of the non-feature and feature losses was heuristically set to 1:0.1 and 1:0.01 respectively to reconstruct visually realistic gestures\footnote{We also attempted to calculate features on a rolling basis for a timeseries and compare two "feature timeseries".
This approach resulted in noisy reconstructions and took prohibitively long to run, so we did not investigate it further.}.

The same na\"ive autoencoder (with architecture detailed in Section~\ref{sec:gen_architecture}) was trained on the WatchAuth gesture data using different loss functions. Each model was trained until convergence with early stopping on its validation error, using the same random 20\% validation split for each choice of loss.

The quality of reconstructions was evaluated quantitatively by the TSTR method (detailed in Section~\ref{sec:tstr}). 

We began by using TSTR for classification of timeseries data into gesture and non-gesture classes, to investigate whether synthetic gestures were reliably distinguishable from non-gesture data. 240 synthetic gestures were generated as the direct reconstructions of randomly selected gestures for the positive class, with 240 non-gesture samples used as the negative class.

\paraHeading{Preliminary results}
Table~\ref{tbl:reconstruction_results} summarises the AUROC and EER values computed for our TSTR gesture recognition task.

We firstly note that models using all loss functions except Soft-DTW achieved strong results on both TSTR tasks. The Soft-DTW model generated samples that achieved superior gesture recognition ability to the original curves on the \textit{RF100} TSTR task but completely failed the \textit{Conv+GRU} TSTR task.

\begin{figure}[t]
	\centering
	\includegraphics[width=0.9\linewidth]{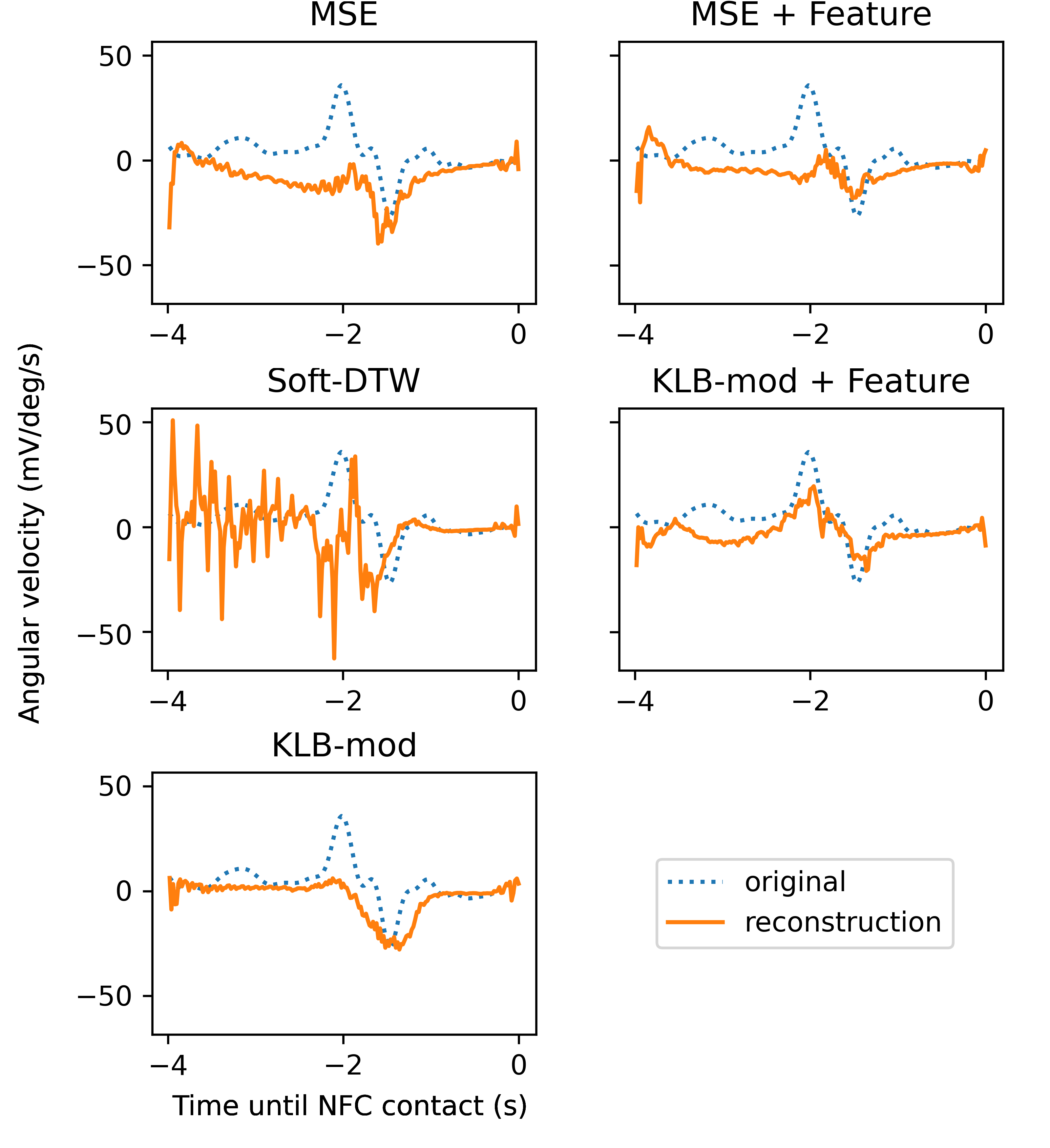}
	\caption{\small Visualising reconstructed gestures along the gyroscope z-axis for autoencoders trained with different loss functions.}
	\label{fig:gyrzlbkeoghstats}
\end{figure}

\begin{figure}[t!]
	\centering
	\includegraphics[width=0.9\linewidth]{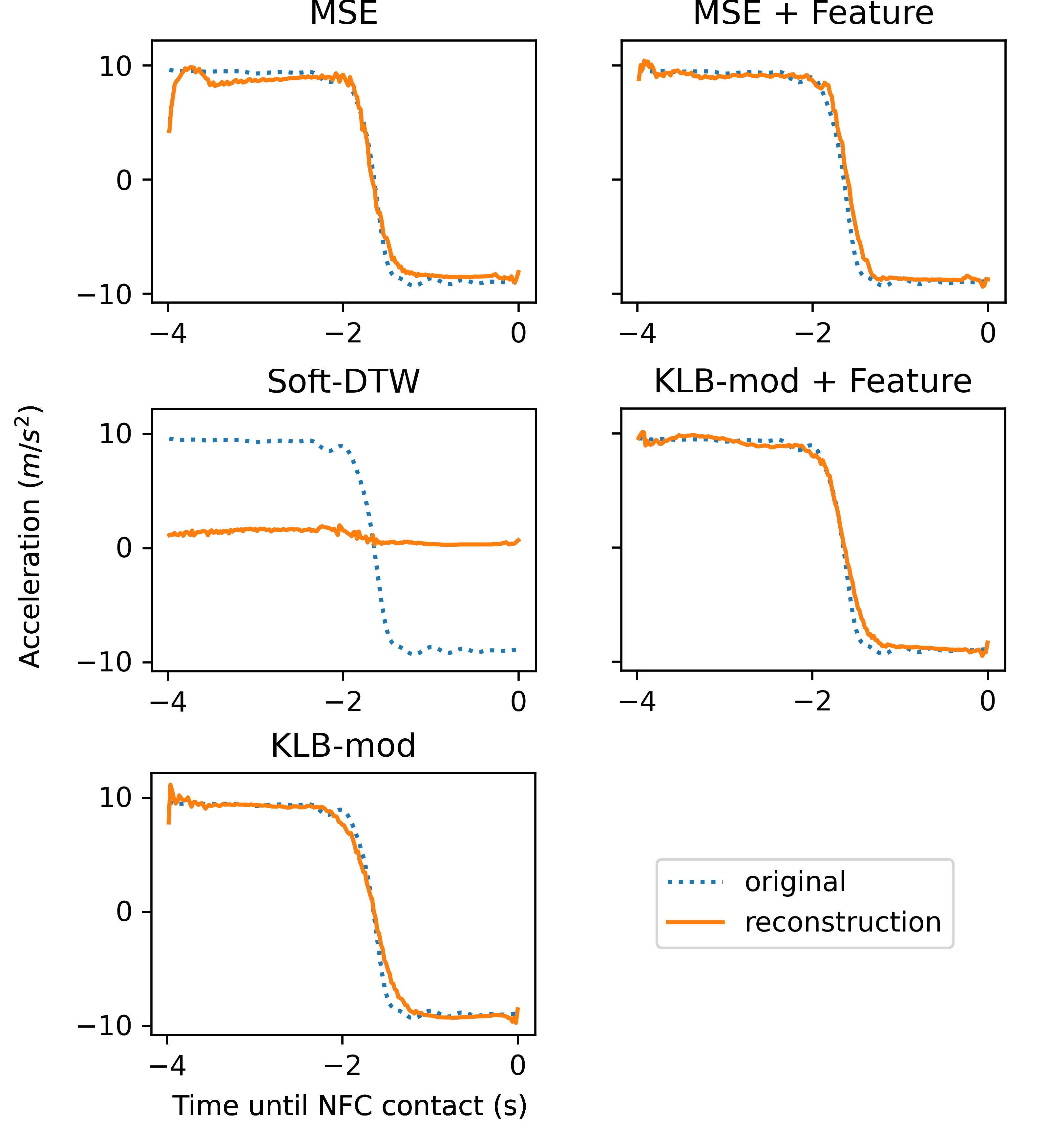}
	\caption{\small Visualising reconstructed gestures along the accelerometer z-axis for autoencoders trained with different loss functions.}
	\label{fig:acczlbkeoghstats}
\end{figure}

Visualising reconstructed gestures gives an explanation. We see in Figure \ref{fig:gyrzlbkeoghstats} that the reconstructed curve from the model using Soft-DTW is very noisy. This could be a result of difficulty in training effectively --- gradients from the loss must be backpropagated through a dynamic programming algorithm, resulting in a highly complex optimisation landscape. In Figure \ref{fig:acczlbkeoghstats} we see a further visualisation of a different channel, with all loss functions except Soft-DTW achieving near-perfect reconstruction.

These reconstructed gestures are clearly extremely different to non-gesture data; perhaps this extreme difference to non-gesture data has resulted in the strong performance on the \textit{RF100} TSTR task despite the unrealistic generated gestures.

\begin{table}[b]
	\caption[]{\small TSTR results for classification between gesture and non-gesture data for autoencoders trained with different choices of reconstruction loss, using both \textit{Conv+GRU} and \textit{RF100} as classifiers.} 
	\label{tbl:reconstruction_results}
	\vskip 0.15in
	\begin{center}
		\begin{sc}
			\input{tables/loss_reconstruction_summary}
		\end{sc}
		\vskip -0.4in
	\end{center}
\end{table}

\section{Autoencoder Regularisation Schema}\label{app:WAE}

\subsection{Wasserstein Autoencoders}

Wasserstein autoencoders (WAEs), proposed by Tolstikhin et al.~\cite{tolstikhinWassersteinAutoEncoders2018}, are an alternative method for regularising an autoencoder. Instead of modelling points as distributions in the latent space and regularising these individual distributions, WAEs use a regularisation term that encourages the distribution of all latent space embeddings to match a Gaussian.
Given $ n $ training points $ \{x_1, ..., x_n\} $, and $ n $ points $ \{ z_1, ..., z_n \} $ sampled independently from a standard Gaussian, we calculate the following regularisation loss:
\begin{align*}
	\ell_{WAE\_regularisation} = \frac{1}{n(n-1)} & \sum_{i \ne j} (E(x_i) - E(x_j))^2 \\ 
 &+ \frac{1}{n(n-1)} \sum_{i \ne j} (z_i - z_j)^2 \\ &- \frac{2}{n^2} \sum_{i,j} (E(x_i) - z_j)^2
\end{align*}

Tolstikhin et al. claim that WAEs retain the favourable qualities of VAEs including stable training and nice latent manifold structure while achieving higher quality reconstructions (which they demonstrate empirically for image data).

\subsection{VAE vs WAE} 

We trialled two approaches for regularising an autoencoder, namely the variational (VAE) and Wasserstein autoencoders (WAE).

Both methods introduce a regularisation term into the autoencoder loss function:
\[ \qquad \ell_{VAE} = \ell_{reconstruction} + \beta \cdot \ell_{regularisation} \]

The hyperparameter $ \beta $ controls the level of regularisation. We exponentially varied the value of $\beta$ between 1 and $ 1e-6 $ and trained a VAE and WAE on the training dataset for each $ \beta $.

We repeated the preliminary TSTR experiments of Appendix \ref{app:losses} using \textit{Conv+GRU} with the VAE and WAE models to investigate whether adding regularisation affected the ability to reconstruct realistic gestures. These results are summarised in Table \ref{tbl:TSTR_intent_vaes_waes}.

\begin{figure}[t!]
	\centering
	\includegraphics[width=0.8\linewidth]{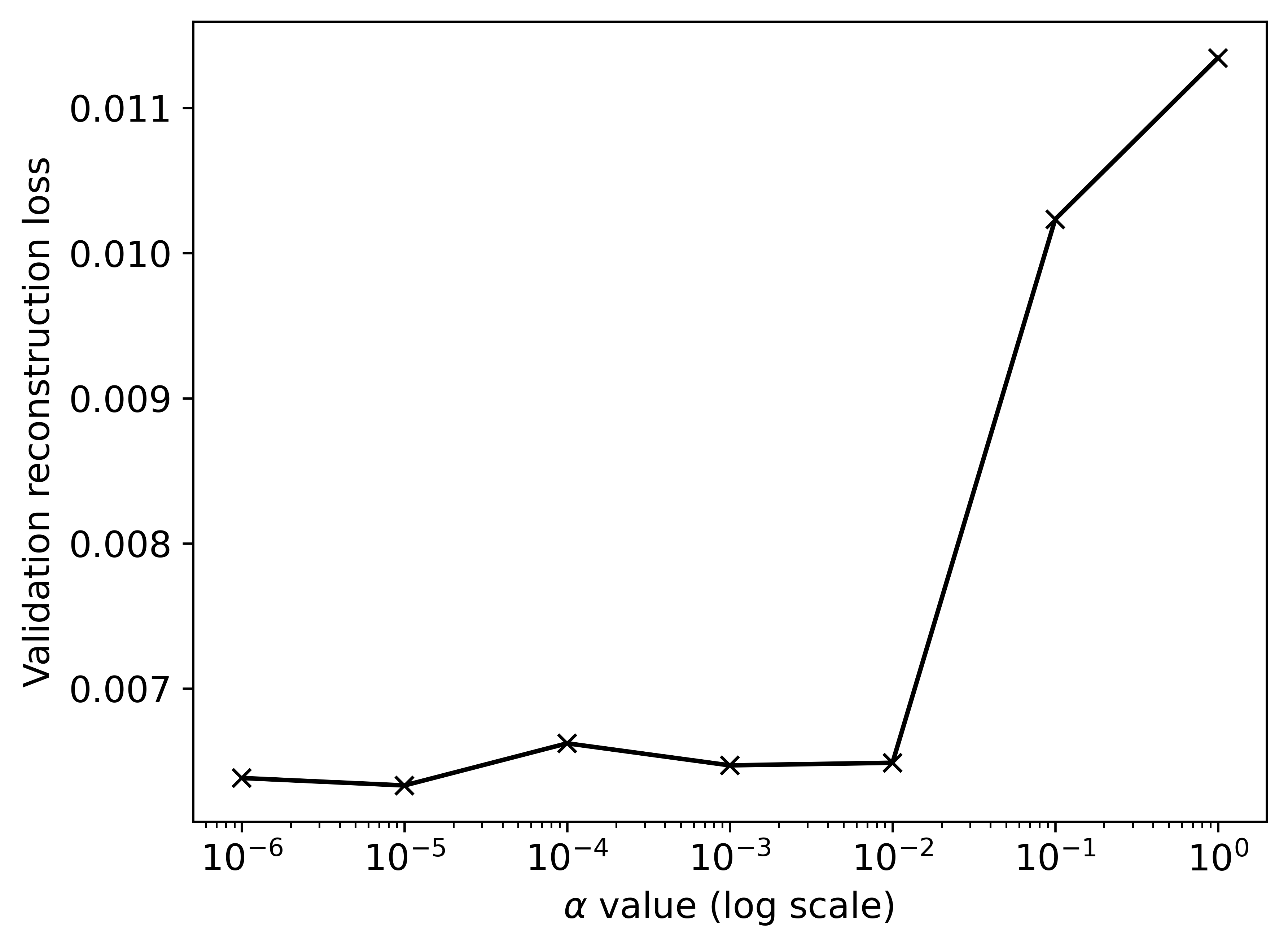}
	\caption{\small Validation reconstruction loss for VAE-based models trained with varying values of $\alpha$, showing the effect of additional weighting on the authentication loss in our modified VAE loss function.}
	\label{fig:valreconlossauthvaes}
\end{figure}

\begin{figure}[t]
	\centering
	\includegraphics[width=0.64\linewidth]{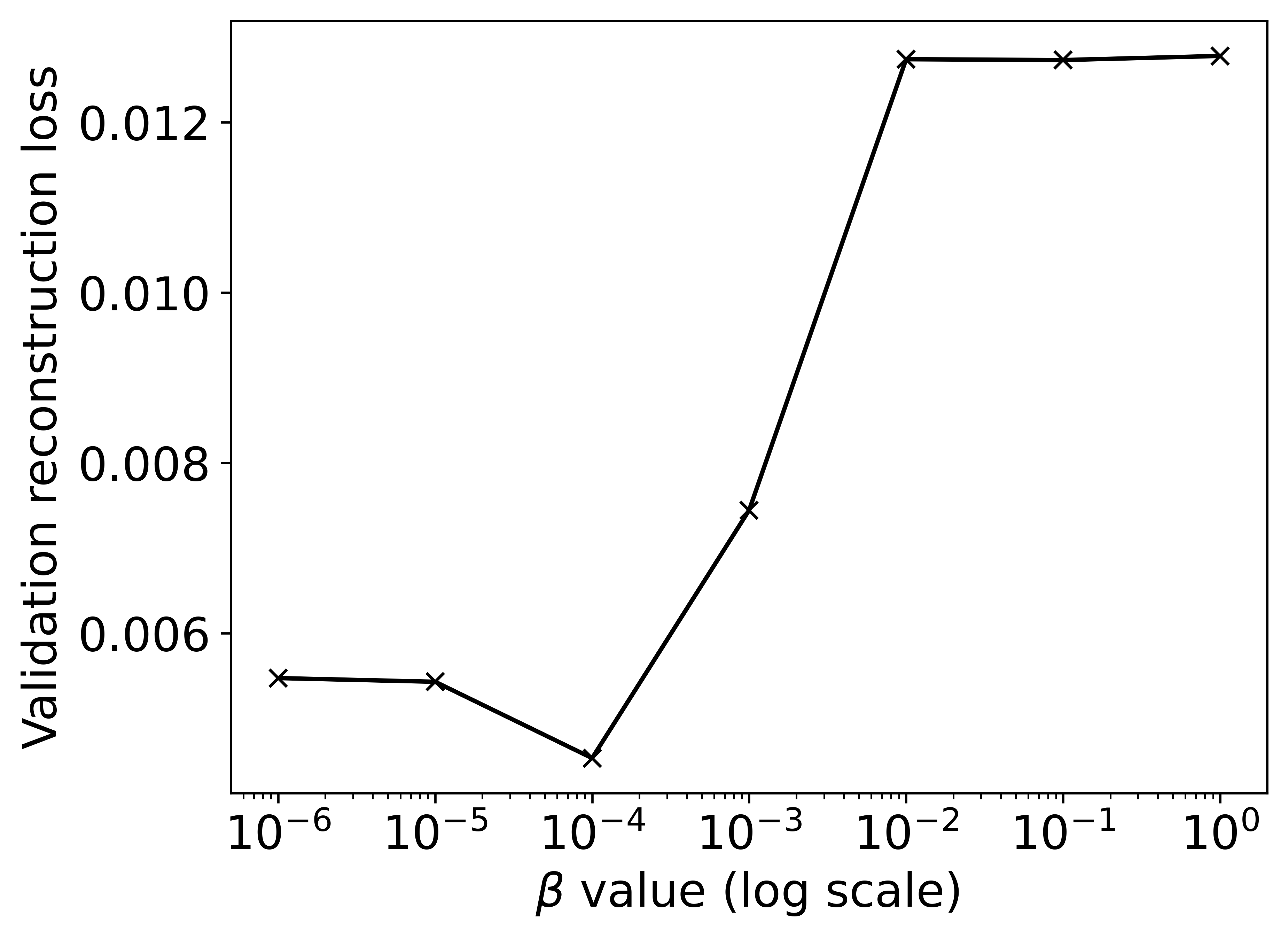}
	\caption{\small Validation reconstruction losses for VAEs trained with varying $ \beta $, showing the effect of additional weighting on the
KL (regularisation) loss in the VAE loss function}
	\label{fig:waesvallosses}
\end{figure}

We can see that the regularisation had a slight effect, but that realistic gestures are still being reconstructed. WAEs performed worse than VAEs, so we proceeded with the VAE regularisation schema.

\begin{table}[t]
	\caption[]{\small TSTR gesture recognition results using \textit{Conv+GRU}, shown for different regularised autoencoders.} 
	\label{tbl:TSTR_intent_vaes_waes}
	\vskip 0.15in
	\begin{center}
		\begin{sc}
			\input{tables/TSTR_intent_vaes_waes}
		\end{sc}
		\vskip -0.4in
	\end{center}
\end{table}

\section{Effect of authentication loss}\label{app:auth_loss}

Figure~\ref{fig:valreconlossauthvaes} shows reconstruction losses for different values of $\alpha$. Figure~\ref{fig:waesvallosses} is discussed in Section~\ref{sec:autoencoder_regularisation}.

\end{document}

%% file: tables/TSTR_auth.tex
\begin{tabularx}{\linewidth}{l|YYY}
	\toprule
	Sampling Strategy & AUROC & EER interval & FAR@0 \\
	\midrule
	Original data only & 0.83 & (0.14, 0.28) & 1.00 \\
	\midrule
	Neighbourhood     & 0.83 & (0.23,0.26) & 0.79 \\
	Self-mixed        & 0.82 & (0.23,0.27) & 0.78 \\
	Adversarial       & 0.84 & (0.21,0.25) & 0.74 \\
	Same-user         & 0.82 & (0.23,0.27) & 0.77 \\
	\bottomrule
\end{tabularx}

%% file: tables/loss_reconstruction_summary.tex
\begin{tabularx}{\linewidth}{l|YYYY}
	\toprule
	& \multicolumn{2}{c}{Conv+GRU} & \multicolumn{2}{c}{RF100}\\
	Loss function       & AUROC & EER & AUROC & EER \\
	\midrule
	No reconstruction   & 0.94 & 0.14 & 0.97 & 0.054  \\
	\midrule
	MSE                 & 0.88 & 0.19 & 0.97 & 0.054  \\
	Soft-DTW            & 0.30 & 0.65 & 0.99 & 0.033  \\
	KLB-mod             & 0.88 & 0.18 & 0.97 &  0.056 \\
	MSE + Feature       & 0.90 & 0.18 & 0.97 & 0.055  \\
	KLB-mod + Feature & 0.92 & 0.16 & 0.97 &  0.056 \\
	\bottomrule
\end{tabularx}

%% file: tables/TSTR_intent_vaes_waes.tex
\begin{tabularx}{\linewidth}{l|YY}
	\toprule
	Regularisation    & AUROC & EER \\
	\midrule
	Unregularised     & 0.92 & 0.16 \\
        \midrule
	VAE               & 0.90 & 0.17 \\
	WAE               & 0.87 & 0.20 \\
	\bottomrule
\end{tabularx}